\documentclass[fleqn,usenatbib,useAMS]{mnras}

\usepackage{hyperref}

\usepackage[T1]{fontenc}
\usepackage{ae,aecompl}

\usepackage{graphicx}	% Including figure files
\usepackage{amsmath}	% Advanced maths commands
\usepackage{amssymb}	% Extra maths symbols

\title[Planets orbiting PH-2 and Kepler-103]{Using HARPS-N to characterise the long-period planets in the PH-2 and Kepler-103 systems}

\author[S.C. Dubber et al.]{
Sophie C. Dubber$^{1,2}$\thanks{E-mail: dubber@roe.ac.uk},
Annelies Mortier$^{3,2}$,
Ken Rice$^{1,4}$,
Chantanelle Nava$^{5}$,
\newauthor
Luca Malavolta$^{6}$,
Helen Giles$^{7}$,
Adrien Coffinet$^{7}$,
David Charbonneau$^{5}$,
\newauthor
Andrew Vanderburg$^{8,\dagger}$,
Aldo S. Bonomo$^{9}$,
Walter Boschin$^{10,11,12}$,
Lars A. Buchhave$^{13}$,
\newauthor
Andrew Collier Cameron$^{2}$,
Rosario Cosentino$^{10}$,
Xavier Dumusque$^{7}$,
\newauthor
Adriano Ghedina$^{10}$,
Avet Harutyunyan$^{10}$,
Rapha\"elle D. Haywood$^{5,\dagger}$,
David Latham$^{5}$,
\newauthor
Mercedes L\'opez-Morales$^{5}$,
Giusi Micela$^{14}$,
Emilio Molinari$^{15}$,
Francesco A. Pepe$^{7}$,
\newauthor
David Phillips$^{5}$,
Giampaolo Piotto$^{16}$,
Ennio Poretti$^{10,17}$,
Dimitar Sasselov$^{5}$,
\newauthor
Alessandro Sozzetti$^{9}$,
St\'ephane Udry$^{7}$
\\
$^{1}$SUPA, Institute for Astronomy, Royal Observatory, University of Edinburgh, Blackford Hill, Edinburgh EH93HJ, UK\\
$^{2}$School of Physics and Astronomy, University of St Andrews, North Haugh, St Andrews, Fife, KY16 9SS\\
$^{3}$Astrophysics Group, Cavendish Laboratory, University of Cambridge, JJ Thomson Avenue, Cambridge CB3 0HE, UK\\
$^{4}$Centre for Exoplanet Science,  University of Edinburgh,  Edinburgh,  UK\\
$^{5}$Center for Astrophysics | Harvard \& Smithsonian, 60 Garden Street, Cambridge, MA 02138 USA\\
$^{6}$INAF - Osservatorio  Astrofisico  di  Catania,  Via  S.  Sofia  78,  I-95123, Catania, Italy\\
$^{7}$Observatoire Astronomique de l'Universit\'e de Gen\`eve, Chemin des Maillettes 51, Sauverny, CH-1290, Switzerland\\
$^{8}$Department of Astronomy, The University of Texas at Austin, 2515 Speedway, Stop C1400, Austin, TX 78712, USA\\
$^{9}$INAF - Osservatorio Astrofisico di Torino, via Osservatorio 20, 10025 Pino Torinese, Italy\\
$^{10}$INAF - Fundaci\'on Galileo Galilei, Rambla Jos\'e Ana Fernandez P\'erez 7, E-38712 Bre\~na Baja, Tenerife, Spain\\
$^{11}$Instituto de Astrof\`isica de Canarias (IAC), Calle V\'ia L\'actea s/n, 38205 La Laguna, Tenerife, Spain\\
$^{12}$Departamento de Astrof\`isica, Universidad de La Laguna (ULL), 38206 La Laguna, Tenerife, Spain\\
$^{13}$DTU Space, National Space Institute, Technical University of Denmark, Elektrovej 328, DK-2800 Kgs. Lyngby, Denmark\\
$^{14}$INAF - Osservatorio Astronomico di Palermo, Piazza del Parlamento 1, I-90134 Palermo, Italy\\
$^{15}$INAF - Osservatorio Astronomico di Cagliari, via della Scienza 5, 09047, Selargius, Italy\\
$^{16}$Dip. di Fisica e Astronomia Galileo Galilei - Universit\`a di Padova, Vicolo dell'Osservatorio 2, 35122, Padova, Italy\\
$^{17}$INAF - Osservatorio Astronomico di Brera, Via E. Bianchi 46, 23807 Merate (LC), Italy\\
$\dagger$NASA Sagan Fellow\\
\\
\\
\\
\\
\\
\\
\\
\\
\\
\\
\\
\\
\\
\\
\\
\\
\\
\\
\\
\\
\\
}

\date{Accepted XXX. Received YYY; in original form ZZZ}

\pubyear{2019}

\begin{document}
\label{firstpage}
\pagerange{\pageref{firstpage}--\pageref{lastpage}}
\maketitle

\newpage

\begin{abstract}
We present confirmation of the planetary nature of PH-2b, as well as the first mass estimates for the two planets in the Kepler-103 system. PH-2b and Kepler-103c are both long-period and transiting, a sparsely-populated category of exoplanet. We use {\it Kepler} light-curve data to estimate a radius, and then use HARPS-N radial velocities to determine the semi-amplitude of the stellar reflex motion and, hence, the planet mass.  For PH-2b we recover a 3.5-$\sigma$ mass estimate of $M_p = 109^{+30}_{-32}$ M$_\oplus$ and a radius of $R_p = 9.49\pm0.16$ R$_\oplus$.  This means that PH-2b has a Saturn-like bulk density and is the only planet of this type with an orbital period $P > 200$ days that orbits a single star.  We find that Kepler-103b has a mass of $M_{\text{p,b}} = 11.7^{+4.31}_{-4.72}$ M$_{\oplus}$ and Kepler-103c has a mass of $M_{\text{p,c}} = 58.5^{+11.2}_{-11.4}$ M$_{\oplus}$. These are 2.5$\sigma$ and 5$\sigma$ results, respectively. With radii of $R_{\text{p,b}} = 3.49^{+0.06}_{-0.05}$ R$_\oplus$, and $R_{\text{p,c}} = 5.45^{+0.18}_{-0.17}$ R$_\oplus$, these results suggest that Kepler-103b has a Neptune-like density, while Kepler-103c is one of the highest density planets with a period $P > 100$ days.  By providing high-precision estimates for the masses of the long-period, intermediate-mass planets PH-2b and Kepler-103c, we increase the sample of long-period planets with known masses and radii, which will improve our understanding of the mass-radius relation across the full range of exoplanet masses and radii.  
\end{abstract} 

\begin{keywords}
planets and satellites: composition -- techniques: photometric -- techniques: radial velocities -- techniques: spectroscopic
\end{keywords}

%%%%%%%%%%%%%%%%%%%%%%%%%%%%%%%%%%%%%%%%%%%%%%%%%%

%%%%%%%%%%%%%%%%% BODY OF PAPER %%%%%%%%%%%%%%%%%%

\section{Introduction} \label{sec:intro}

The photometric transit method, which allows for an estimate of a planet's radius \citep[e.g.][]{charbonneau00,batalha11}, and the radial velocity (RV) method, which allows for an estimate of a planet's mass \citep[e.g.][]{bonfils11,vogt10} can be combined to infer a planet's internal composition and determine if it still retains a volatile envelope \citep[e.g.,][]{rogers15}. Ideally, we would like  a large sample of exoplanets, ranging from small, rocky planets up to large gas giants, with a wide range of orbital periods and precise mass and radius estimates, so that we can develop a good understanding of the potential underlying mass-radius (MR) relation.

The advent of NASA's {\it Kepler} spacecraft has significantly increased the sample of exoplanets with precise radius estimates.  Highly precise radial velocity spectrometers, such as HARPS \citep{mayor03} and HARPS-N \citep{consentino12}, have also allowed for follow-up observations to determine precise mass estimates, especially for small planets. For example, a specific goal of the HARPS-N collaboration has been to follow-up and provide precise mass estimates for some of the small exoplanets discovered by {\it Kepler} and {\it K2}. Examples include Kepler-93b \citep{dressing15}, Kepler-21b \citep{lopezmorales16} and K2-263b \citep{mortier18}.  

Of particular current interest is the detection of a gap in the radius distribution for lower-mass planets \citep{fulton17}.  It appears that planets either have radii less than $\sim 1.5$ R$_\oplus$ and are pre-dominantly rocky, or have radii above $\sim 2$ R$_\oplus$ and still retain a substantial volatile atmosphere. Although the origin of this gap is not clear, the existence of such a gap was predicted \citep{owen13, lopez14} for close-in planets subject to high levels of irradiation.  The population of close-in, low-mass exoplanets appears consistent with this photo-evaporation scenario \citep[e.g.][]{vaneylen18,rice19}.

This does mean that less attention has been paid to slightly larger planets, of intermediate mass, that may help us to better understand the overall relation between planet mass and radius. Obtaining highly precise mass measurements can require committing a lot of time to the observations. The number of observations required scales as $N \propto (\sigma/K)^{2}$, where $\sigma$ is the precision of a single RV measurement, and $K$ is RV semi-amplitude \citep{gaudi07}. Clearly, the number of observations needed to obtain a highly precise mass can vary for each target and set of observing conditions, but in general for a given error level, more measurements are required for a planet of a lower mass. This focus on low mass planets means that we have spent less time observing those systems with known planets that are probably of intermediate mass. 

There is also a scarcity of long-period planets (here defined as $P>100$\, days) with both masses and radii precisely determined. This is primarily due to the low probability of observing the transit of such a planet, which scales inversely with the orbital period \citep{beatty08}. This gap in the population is thus a result of observational biases, rather than a real feature of the distribution of exoplanets. Radial velocity observations of long-period, transiting planets are also rare, since longer observation times are needed to cover enough of an orbit to constrain the radial velocity curve. Despite the observational challenges, planets in this part of the parameter space deserve to be extensively studied, which will also remove the manifestation of observational biases from our understanding of the population. Currently, only a handful of long-period planets are well characterised: examples include HD80606b \citep{naef01,moutou09} characterised by extensive RV monitoring, and Kepler-51 \citep{ford11,hadden17} characterised by transit timing variations (TTVs). This population of giant planets is a key tool in testing predictions of internal structure and evolution of giant planets. For example, their increased orbital distance means they ought to be significantly less inflated than their closer-in siblings. Furthermore, these detections are crucial for filling out the period-density parameter space: direct imaging observations regularly detect very long-period planets, but determining a density for these objects is challenging. 

There have been a number of attempts to develop a mass-radius relation, which would increase the number of exoplanets with density estimates. \citet{seager07} suggested a power law function for solid exoplanets, allowing the slope to change depending on the mass range. They used simplified equations of state for iron cores and silicate- or water-dominated mantles. \citet{lissauer11} fitted a relation to mass and radius data for Solar System planets, but recognised that this simpler approach was insufficient when considering large planets. However, a unique MR relation may not exist at all \citep[see][for an overview]{ning2018}. In 2014, \citeauthor*{weissmarcy14} used a larger sample of 65 Earth-to-Neptune sized exoplanets, and found that a simple power-law fit had a large scatter in mass for a given radius, due to different compositions. More recently, \citet{wolfgang16} first quantified how this intrinsic scatter varies as a function of radius. \citet{chen17} developed a predictive model using a probabilistic MR relation, that spans a larger parameter space than previous works, and also treats the transition points between different planet populations as inferred parameters. They demonstrate the need for different power laws for the different populations, with clear transitions between each. 

The current dataset is still too limited to prove the existence of an MR relation definitively. Thus it has become increasingly important to use all available observational data to increase the sample of planets with known masses, including those where the result may be less precise than we would like. Typically, detection papers focus on $>$ 6$\sigma$ results for planet masses \citep[e.g.][]{haywood18,malavolta18}, and even higher precision for radii. Requiring this level of precision for detections places severe practical limits on the number of planets for which well-constrained masses and radii are available. However, detections of lower significance can still add useful data points to the sample of known planets, as long as well-constrained posterior mass distributions are produced.

When HARPS-N was initially commissioned, one of the priorities of the Science Team was to perform follow up investigations of objects identified by the {\it Kepler} survey \citep{borucki10}. Subsequently this became follow-up for the K2 survey and it has been further extended by the launch of the Transiting Exoplanet Survey Satellite (TESS) \citep{ricker15}, for which HARPS-N will also aim to carry out follow-up observations. Objects observed in the K2 survey were generally brighter targets, meaning that some of the {\it Kepler} objects initially observed by HARPS-N were later classed as lower priority. Even the best techniques at the time were unable to observe any clear planetary signals. However, the progression of analysis techniques in the six years since the beginning of HARPS-N observations now means that these lower priority targets may in fact still be conducive to well-constrained results. We chose to revisit some of these targets, to check that good data was not being wasted, and also to potentially increase the sample of exoplanets in a region of parameter space that is currently not well populated. We found that with the improved techniques such as Gaussian process (GP) regression modelling \citep{haywood14}, we were indeed able to use this revisited data to find, in some cases, well-constrained planetary masses.

Here, we revisit two HARPS-N targets, PH-2 and Kepler-103 (see Table \ref{tab:stell_par} for stellar parameters), identified as planetary candidates by the {\it Kepler} survey \citep{borucki10}. Both PH-2 and Kepler-103 are solar-type stars, with G3 and G2 spectral types and visual magnitudes of $V=12.70$ and $V=12.36$, respectively \citep{henden15}. We performed multiple follow-up RV observations using HARPS-N, aiming to confirm the planetary nature of the candidates and to provide mass estimates.  We report a 3.5$\sigma$ RV semi-amplitude value for the Saturn-like, long period planet PH-2b (also known as Kepler-86b). This planet was first identified as a planetary candidate (KOI-3663b) by \citet{wang13}. We were also able to constrain the masses of the two planets in the Kepler-103 system (KOI-108), presenting a 2.5$\sigma$ result for Kepler-103b and a 5$\sigma$ result for Kepler-103c.  Both Kepler-103b and Kepler-103c do have previously published mass constraints \citep{marcy14}, but neither mass was precisely measured in this earlier work. We find Kepler-103b to be a super Earth on a relatively close-in orbit, whereas Kepler-103c is a wide orbit, high density giant planet. 

We present the {\it Kepler} and HARPS-N observations for each of the systems in Section \ref{sec:obs}. We discuss the stellar analysis in Section \ref{sec:stellar}. The transit and RV analysis are descried in Sections \ref{sec:transit} and \ref{sec:RV_analysis}, respectively. We discuss the results in Section \ref{sec:disc}, placing these measurements in the context of the existing field. Finally, we conclude in Section \ref{sec:concl}.

%%%%%%%%%%%%%%%%
% Observations %
%%%%%%%%%%%%%%%%

\section{Observations} \label{sec:obs}

\subsection{Kepler Photometry} \label{sec:obs_phot}

PH-2 was monitored with {\it Kepler} in long-cadence mode, observing every 29.4 minutes, in all quarters Q0-Q17, with transits only observed in Q4,Q7,Q10,Q13 and Q16. These observations cover a total time period of 1470 days (BJD 2454953.0375 - 2456423.500694).

Kepler-103 was also monitored in long-cadence mode in quarters Q0-Q17, covering the same time period. Transits were uncovered in every quarter of the data.

Long-term variations (on a {\it Kepler} quarter timescale) are present in the simple aperture flux (SAP), as a result of differential velocity aberration. If this is not removed, it can obscure stellar rotation signals. We have chosen to work with the Presearch Data Conditioning SAP light-curve from DR21 \citep{smith2012,stumpe2012,stumpe2014} as this version more effectively removes these trends (see \citet{haywood18} for further details). The full {\it Kepler} light-curves for both stars are shown in Figures \ref{fig:PH2_full_lightcurve} and \ref{fig:K103_full_lightcurve}.

We followed a procedure similar to \citet{haywood18}. In short, we fit the PDCSAP short-cadence light-curves produced by the {\it Kepler} pipeline \citep{smith2012,stumpe2012}. We flatten the light-curve by fitting polynomials to the out-of-transit light-curves near transits, and dividing by the best-fit polynomial. Depending on the number of points in the out-of-transit region near the transits, we chose either first order (linear) or second order (parabola) polynomials. We exclude outliers from the phase-folded light-curve by dividing it into bins of a few minutes. These outliers are especially prevalent in time periods towards the end of the original {\it Kepler} mission, during which the second of four reaction wheels was close to failing. We then exclude any 3-sigma outliers that lie in each of these bins. 
\subsection{HARPS-N Spectroscopy} \label{sec:obs_spec}

To collect spectra of our targets, we used the high-resolution HARPS-N spectrograph (R = 115000) which is installed on the 3.6-m Telescopio Nazionale Galileo (TNG) at the Observatorio del Roque de los Muchachos in La Palma, Spain \citep{consentino12,cosentino14}. Observations were taken as part of the HARPS-N Collaboration's Guaranteed Time Observation's (GTO) programme. The spectra were used for both stellar parameter determination and for obtaining high-precision radial velocities (RVs).

The spectra were reduced with version 3.9 of the HARPS-N Data Reduction Software (DRS), which includes corrections for color systematics introduced by variations in seeing \citep{consentino12}. The observed spectra are cross-correlated with a spectral template chosen to be the closest match to the spectral type of the target \citep{pepe02}. In this case, we used a G2 template for both objects.  A Gaussian is fitted to the resulting cross-correlation function (CCF) so that the RV values and the associated errors can be determined. Other properties of the CCF can be used to ascertain activity indicators, such as the full width at half maximum (FWHM), the contrast, and bisector span (BIS). Additionally, the chromospheric indicator $\log R'_{\rm HK}$ was calculated from the Ca H\&K lines \citep{noyes84}.  

\subsubsection{PH-2 (Kepler-86)} \label{sec:obs_PH2}
PH-2 was observed using HARPS-N in regular intervals between 29 September 2013 and 19 May 2016, to observe multiple orbits of the $P = 282$ days planet candidate.  We collected 33 spectra, with typical exposure times of 1800s (the exceptions being those taken on BJD = 2457164.71, 2457164.72, 2457165.70, 2457165.71, when the exposure times were 900s due to exceptional weather conditions).

\begin{figure}
    \centering
    \includegraphics[width=8.7cm]{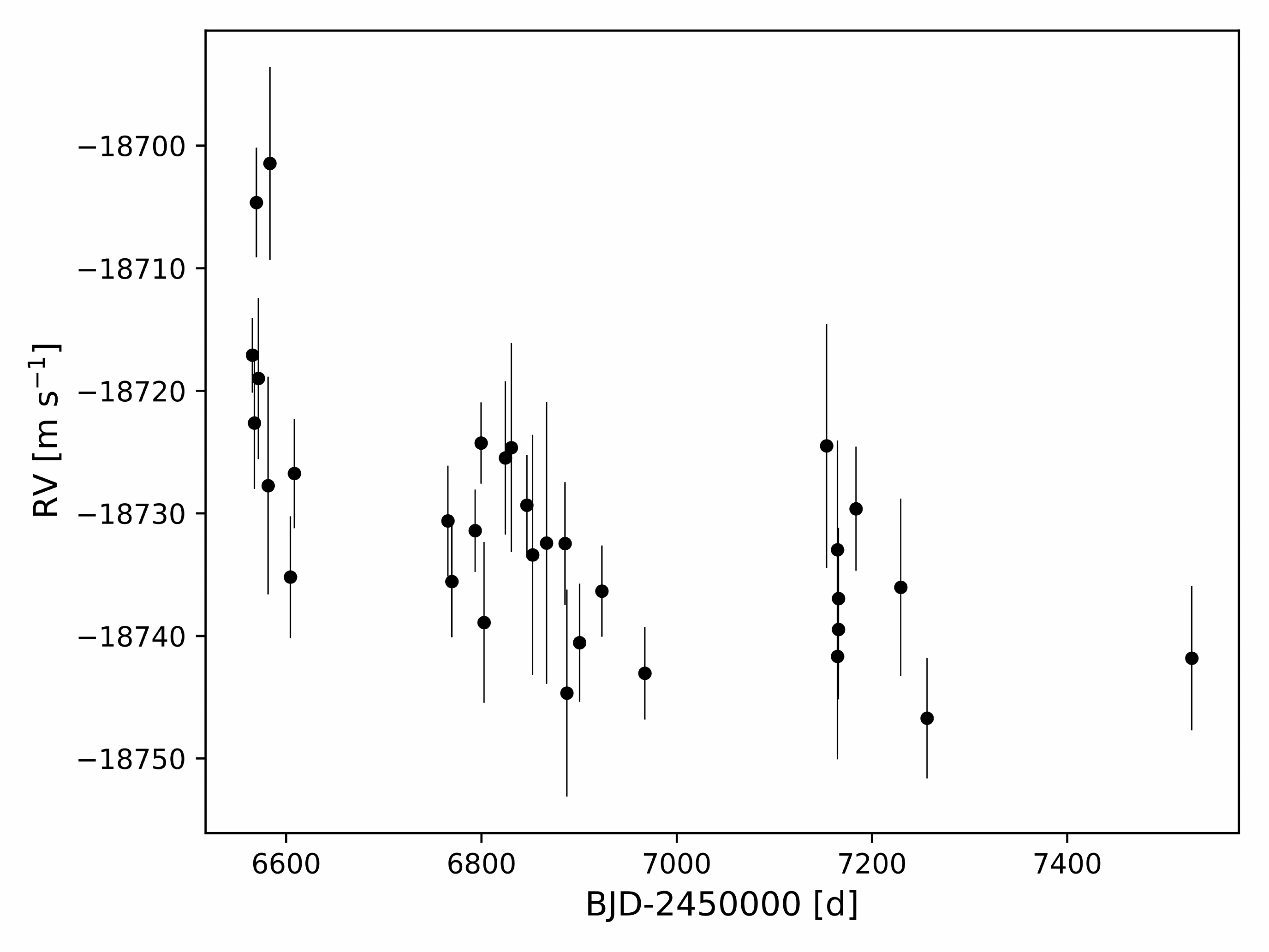}  
    \caption{The HARPS-N RVs for PH-2, plotted against time.}
    \label{fig:RV_PH2}
\end{figure}

The majority of RV data points were obtained by observing with the 2nd fibre on-sky. However, we eliminated one RV that was extracted from a spectrum observed with simultaneous thorium-argon rather than a simultaneous sky fiber (BJD = 2456573.50). Different templates used for observing can create an offset in RV measurements, and in this case the RV was significantly discrepant when including it in the dataset. This led to a final dataset of 32 RVs with signal-to-noise ratios in the range S/N = 13.7 - 38.9 at 550nm (average S/N = 24.6). The average RV internal uncertainty was 6.1 m s$^{-1}$. 
The HARPS-N RVs for PH2 are shown in Figure \ref{fig:RV_PH2}, and the RV data, associated 1$\sigma$ errors, and activity indicators are shown in Table \ref{tab:rv_3663}.

\subsubsection{Kepler-103 (KOI-108)} \label{sec:obs_kep103}
We observed Kepler-103 regularly from 22 May 2014 until 06 October 2015.  We then re-observed it starting on 30 August 2018 and ending on 01 November 2018.  Our strategy was to take one observation per night with an exposure time of 1800s (apart from observations on BJD=2456865.56 and BJD=2456866.51 when the exposure times were 1600s and 1500s respectively) and to observe the system for a few stellar rotation periods. This improved the window function and provided a sampling adequate to pick up potential signals due to stellar activity.

We collected a total of 60 RV spectra, with signal-to-noise ratios in the range S/N = 13.8 - 51.9 at 550nm (average S/N = 31.1), and with an average RV internal uncertainty of 5.1 m s$^{-1}$.

\begin{figure}
    \centering
    \includegraphics[width=8.5cm]{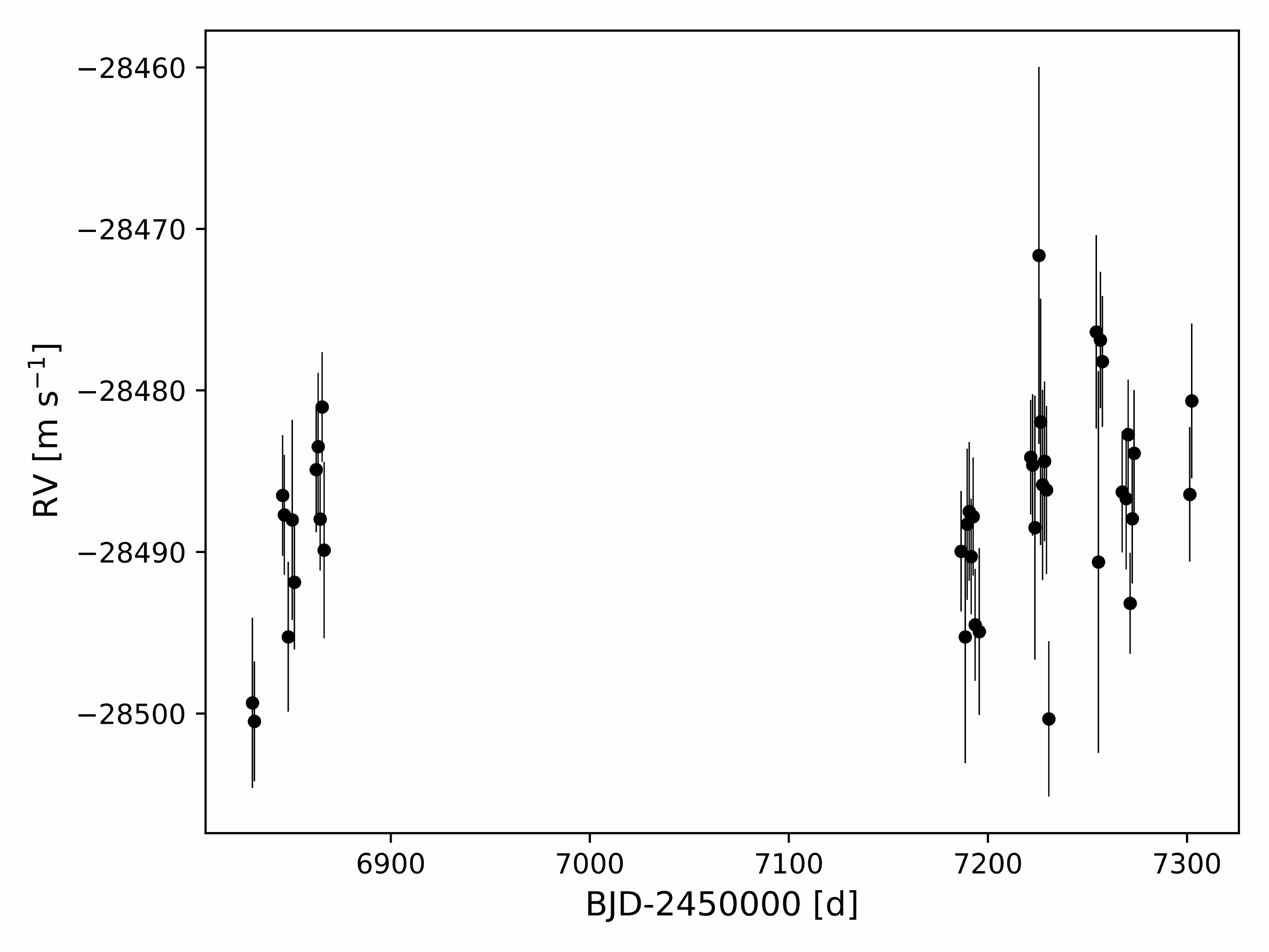}
    \caption{The HARPS-N RVs for Kepler-103, plotted against time.}
    \label{fig:RV_Kep103}
\end{figure}

The HARPS-N RVs for Kepler-103 are shown in Figure \ref{fig:RV_Kep103} and the RV data, associated 1$\sigma$ errors, and activity indicators are shown in Table \ref{tab:rv_108}. 

%%%%%%%%%%%%%%%%%%%%
% Stellar analysis %
%%%%%%%%%%%%%%%%%%%%

\section{Stellar Analysis} \label{sec:stellar}
 
 \subsection{Atmospheric parameters}
 
We used the high-resolution spectra from HARPS-N to determine the stellar atmospheric parameters. Using the estimated RVs, we shifted all spectra to the rest frame and then stacked the spectra, enhancing our signal-to-noise. Equivalent widths (EWs) were determined automatically using {\tt ARESv2}\footnote{Available at \url{http://www.astro.up.pt/~sousasag/ares/}.} \citep{sousa15}, using a line list of roughly 300 neutral and ionised iron lines taken from \citet{sousa11}. 

These EWs were then used as input to {\tt MOOG}\footnote{Available at \url{http://www.as.utexas.edu/~chris/moog.html}.} \citep{sneden73} for line analysis to obtain effective temperatures, metallicity, and surface gravity, under the assumption of local thermodynamic equilibrium (LTE). This process was implemented using {\tt FASMA} \footnote{Available at \url{http://www.iastro.pt/fasma/}}, as described by \citet{andreasen17}, and a correction was applied to the surface gravity in accordance with \citet{mortier14}. Finally, we added systematic errors in quadrature, as outlined in \citet{sousa11}, necessary to account for differences in values found using different methods. For the effective temperature, we added a systematic error of 60 K, for metallicity we added 0.04 dex, and for surface gravity we added 0.1 dex. The resulting parameter values for each star are given in Table \ref{tab:stell_par}.

As outlined in \citet{borsato19}, using multiple methods to find different sets of stellar photospheric parameters is an important tool for getting realistic errors on the stellar mass and radius.
Consequently, we also used {\tt CCFpams} \footnote{Available at \url{https://github.com/LucaMalavolta/CCFpams}} to calculate an independent estimation of the same parameter values \citep{malavolta17b}. The \citet{mortier14} surface gravity correction was again used. The parameter values are consistent with those determined using {\tt FASMA}, with the exception of a 1$\sigma$ difference in the metallicity of PH-2. Strong agreement of two independent methods gave us confidence that using only the {\tt FASMA} results was sufficient for determining the stellar masses and radii.

\begin{table*}
	\centering
	\caption{Stellar parameters for each of the target stars. We show the 2MASS $J$, $H$, and $K$ magnitudes, taken from \citet{skrutskie06}, $V$ magnitudes from APASS DR9 \citep{henden15}, values of $T_{\text{eff}}$, [Fe/H] and log $g$ determined using the {\tt FASMA} EW Method, and the {\it Gaia} parallaxes \citep{gaia16,gaia18}. We also show the average value of log$R'_{HK}$, determined from the HARPS-N spectra, the projected rotational velocity from \citet{petigura17}, the $T_{\rm eff}$, [Fe/H], log $g$ values from the {\tt isochrones} analysis (which agree well with those used as priors) and the resulting mass and radius values in solar units.}
	\label{tab:stell_par}
	\begin{tabular}{llll}
	    \hline
	    Parameter & Descriptor & PH-2 & Kepler-103 \\
	    \hline
	    \noalign{\smallskip}
		$J$ & 2MASS $J$ mag & $11.501 \pm 0.023$ & $11.193 \pm 0.021$ \\
		$H$ & 2MASS $H$ mag & $11.182 \pm 0.030$ & $10.941 \pm 0.016$ \\
		$K$ & 2MASS $K$ mag & $11.116 \pm 0.022$ & $10.873 \pm 0.017$ \\
		\noalign{\medskip}
		$V$ & APASS $V$ mag & $12.699 \pm 0.010$ & $12.360 \pm 0.026$ \\
		\noalign{\medskip}
		$T_{\rm eff}$ (K) & Effective temperature ({\tt FASMA}) & $5691 \pm 67$ & $6009 \pm 64$ \\
		\noalign{\smallskip}
		${\rm [Fe/H]}$ & Metallicity ({\tt FASMA}) & $-0.03 \pm 0.04$ & $0.16 \pm 0.04$ \\
		\noalign{\smallskip}
		log $g$ (cgs) & Surface gravity ({\tt FASMA}) & $4.42 \pm 0.10$ & $4.29 \pm 0.10$ \\
		\noalign{\medskip}
		$\pi_{Gaia}$ (mas) & {\it Gaia} parallax & $2.880 \pm 0.030$ & $1.992 \pm 0.024$ \\
		\noalign{\medskip}
		$T_{\rm eff}$ (K) & Effective temperature ({\tt isochrones}) & $5711^{+60}_{-59}$ & $6047^{+50}_{-67}$ \\
		\noalign{\smallskip}
		${\rm [Fe/H]}$ & Metallicity ({\tt isochrones}) & $-0.03 \pm 0.04$ & $0.15 \pm 0.04$ \\ 
		\noalign{\smallskip}
		log $g$ (cgs) & Surface gravity ({\tt isochrones}) & $4.45 \pm 0.02$ & $4.17 \pm 0.02$ \\
		\noalign{\smallskip}
		$M_*$ (M$_\odot$) & Stellar mass ({\tt isochrones}) & $0.958 \pm 0.034$ & $1.212^{+0.024}_{-0.033}$ \\
		\noalign{\smallskip}
		$R_*$ (R$_{\odot}$) & Stellar radius ({\tt isochrones}) & $0.961^{+0.016}_{-0.015}$ & $1.492^{+0.024}_{-0.022}$ \\
		\noalign{\medskip}
		$v\sin i$ (km/s) & Projected rotational velocity & $2.0 \pm 1.0$ & $3.1 \pm 1.0$ \\		
		\noalign{\medskip}
		$\log R'_{HK}$ & Average value of the $\log R'_{HK}$ activity indicator & $ -4.848\pm0.117$ & $-5.069\pm0.086$ \\
		\\
		\hline
    \end{tabular} 
\end{table*}

\subsection{Stellar mass and radius}

The stellar mass and radius in each case were then found using the {\tt isochrones} python package \citep{morton15}. This uses both the Mesa Isochrones and Stellar Tracks (MIST, \citet{dotter16}) and the Dartmouth Stellar Evolution Database \citep{dotter08}. As priors we use the $T_{\rm eff}$, log $g$, and [Fe/H] values from the {\tt FASMA} analysis, the 2MASS $J$, $H$ and $K$ magnitudes \citep{skrutskie06}, and the {\it Gaia} parallax from Data Release 2 \citep{gaia16,gaia18}. Using only the 2MASS $J$, $H$, and $K$ magnitudes has been shown to be sufficient for estimating the masses and radii of stars of these spectral types \citep{mayo18}.  We used both the MIST and Dartmouth model grids and posterior sampling was performed using {\tt MultiNest} \citep{feroz08,feroz09,feroz13}. 

The final estimate for each parameter is then determined by taking the 15.865th/84.135th percentiles of the combined posterior samples for all sets of stellar parameters.  We recover a stellar mass and radius for PH-2 of $M_\ast = 0.958 \pm 0.034$ M$_\odot$ and $R_\ast = 0.961^{+0.016}_{-0.015}$ R$_\odot$.  For Kepler-103, we find $M_* = 1.212^{+0.024}_{-0.033}$ M$_\odot$ and $R_* = 1.492^{+0.023}_{-0.022}$ R$_\odot$. Table \ref{tab:stell_par} shows all the resulting parameters from the {\tt isochrones} analysis. The output effective temperature, metallicity, and surface gravity all agree well with both of the methods used to initially estimate these parameters, implying that these results are independent of which method is chosen to supply the prior values. 

\subsection{Stellar Activity}
\label{sec:stell_act}

\begin{table}
    \centering
    \caption{Results of the GP stellar activity analysis of the {\it Kepler} light-curve.}
    \begin{tabular}{lll}
        \hline
	    Parameter & PH-2 & Kepler-103 \\
        \hline
        $P_{\rm rot,ACF}$ & $22.6$\,d & $20.8$\,d \\
		\noalign{\medskip}
        $\theta = P_{\rm rot,GP}$ & $17.16^{+6.14}_{-6.03}$\,d & $21.45^{+1.15}_{-5.90}$\,d \\
		\noalign{\medskip}
        $\lambda$ & $6.29^{+5.96}_{-3.22}$\,d & $11.42^{+2.51}_{-5.32}$\,d \\
		\noalign{\medskip}
        $w$ & $0.286^{+0.225}_{-0.089}$ & $0.242^{+0.102}_{-0.029}$ \\
		\noalign{\medskip}
        $h$ & $0.00055^{+0.00005}_{-0.00004}$\,mag & $0.000085^{+0.000008}_{-0.000007}$\,mag \\
        \hline
    \end{tabular}
    \label{tab:lc_gp}
\end{table}

To assess the level of stellar activity, we looked at the auto-correlation functions (ACFs) of the {\it Kepler} light-curves where we saw different degrees of significant variable structure. Therefore, it was only possible to make estimates of the stellar rotational periods, $P_{\text{rot}}$, rather than determine both the rotational periods and activity lifetimes as described in \citet{giles17}. Estimates for the stellar rotational period were extracted from the location of the first side lobe of the ACF, found through peak detection by means of a parabola fit. Our analysis recovered a rotation period of $P_{\rm rot} = 22.6$ days for PH-2 and $P_{\rm rot} = 20.8$ days for Kepler-103. Since no full fit of the ACF was performed, as in \citet{giles17}, these are just rough estimates.

Additionally, we used the {\it Kepler} light-curves to carry out a Gaussian process (GP) analysis using a quasi-periodic covariance kernel function \citep{haywood14,grunblatt15,angus18}. This combines a squared exponential and standard periodic kernel \citep{haywood14} and allows us to most accurately include the effects of various stellar properties. The kernel is described by

\begin{equation}
\Sigma_{\text{i,j}} = h^{2} \text{exp} \left[-\frac{\text{sin}^{2} [\pi (t_{\text{i}} - t_{\text{j}}) / \theta]}{2 w^{2}} - \left(\frac{t_{\text{i}} - t_{\text{j}}}{\lambda}\right)^{2}\right],
\label{eq:gp}
\end{equation}

where $t_i$ and $t_j$ are two times of observation and $\theta, w, h, \lambda$ are the `hyper-parameters'. These can be related to various stellar properties. For example, $h$ (mag) is the amplitude of the correlated signal in the light-curve caused by stellar activity, and $\theta$ (days) is equivalent to the rotational period of the star ($\theta = P_{\text{rot}}$). The other two parameters are related to the evolution of the active regions on the surface of the star. $\lambda$ (days) is the decay timescale of the active regions, which is tied to their aperiodic variation, and $w$ is the coherence scale, which is linked to the amount of active regions present at any time. The typical values of these parameters are known to varying degrees. Active-region decay times are on the order of weeks to months \citep{giles17}, and foreshortening and limb darkening restrict $w$ to values on the order of 0.5, allowing no more than 2-3 peaks to develop in the light-curve or RV curve per rotation cycle.

We used {\tt PyORBIT}\footnote{Available at \url{https://github.com/LucaMalavolta/PyORBIT}, version 8.} \citep{malavolta16}, a package for modelling planetary and stellar activity signals (See Section \ref{sec:methods_pyorbit}).  This implements the GP quasi-periodic kernel through the {\tt george} package \citep{ambikasaran15}, optimizes the hyperparameters using the differential evolution code {\tt PyDE}\footnote{Available at \url{https://github.com/hpparvi/PyDE}}, and then provides the optimized hyperparameters as starting values for the affine-invariant ensemble sampler {\tt emcee} \citep{foreman13}.

Since the GP regression typically scales with the third power of the number of datapoints, we could not use the full light-curve.  Instead, we used a sample of the light-curve that would cover many rotation periods.  For Kepler-103, we combined Quarters 11, 12, and 13, while for PH-2 we combined Quarters 15, 16 and 17.  In both cases, we then removed the transits and any data points more than 5$\sigma$ from the mean.  We also binned the light-curve every 10 data points and allowed for different offsets and jitters for each Quarter. 

The results are presented in Table \ref{tab:lc_gp}. The rotation periods recovered from the GP analysis are consistent with the results from the ACF analysis and plausible given their spectral type.

%%%%%%%%%%%
% TRANSIT %
%%%%%%%%%%%

\section{Transit analysis} \label{sec:transit}

\begin{figure*}
    \centering
    \includegraphics[width=\textwidth]{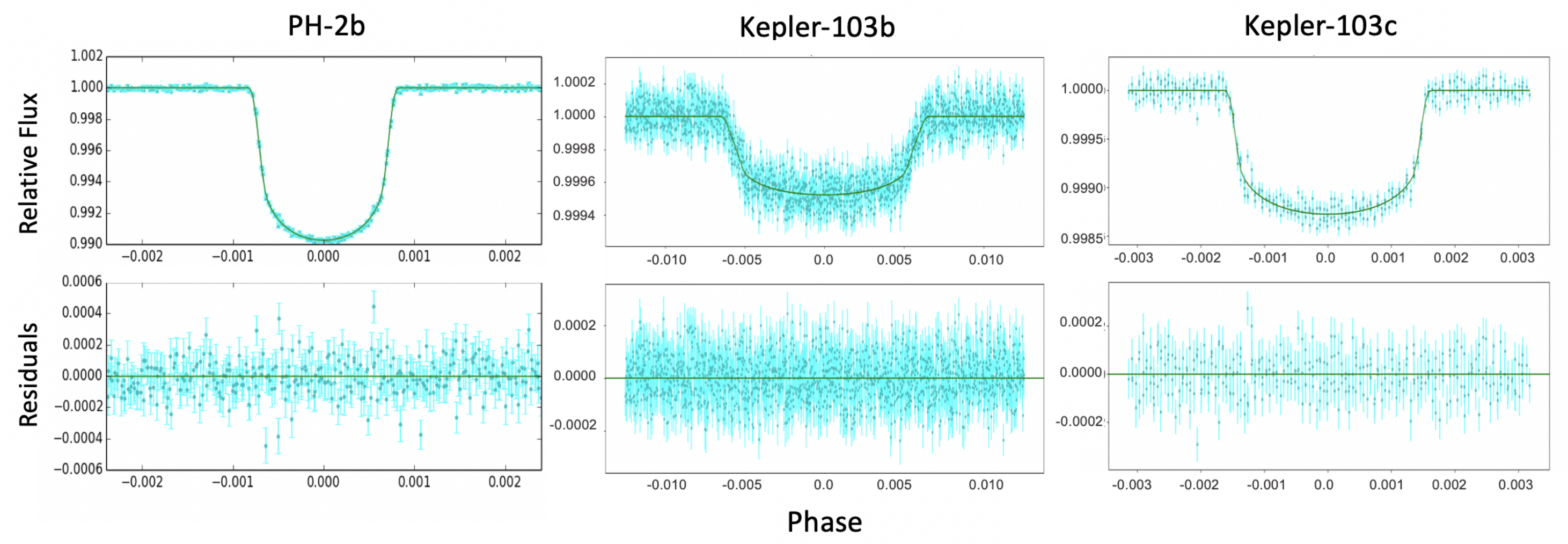}
    \caption{Phase-folded {\it Kepler} light-curve for PH-2b, Kepler-103b and Kepler-103c, demonstrating the transits of the planet candidates in the systems.}
    \label{fig:kepler}
\end{figure*}

\begin{table*}
    \centering
    \caption{Parameters modeled in the \emph{EXOFAST v2} analysis and their prior probability distributions. See text for an explanation of the different prior parameters used for each system. Relevant final values are given in Tables \ref{tab:3663_output} \& \ref{tab:103_output}.}
    \begin{tabular}{lccr}
        \hline
        ID & Parameter & & Prior   \\
        \hline
        PH-2 & Eccentricity & $e$ & Uniform (0.051067, 0.155592) \\
         & Effective Temp. & $T_{\text{eff}}$ & Gaussian (5961, 67) \\
         & Metallicity & [$Fe/H$] & Gaussian (-0.03, 0.04) \\
         &&&\\
        Kepler-103 & Orbital Period & $P_{\text{b}}$ & (15.96532718, 0.000012) \\
         & " & $P_{\text{c}}$ & Gaussian (179.609803, 0.0002) \\
         & Central Transit Time & $T_{\text{C,b}}$ & Gaussian (844.65246, 0.00039) \\
         & " & $T_{\text{C,c}}$ & Gaussian (834.15977, 0.000447) \\
         & Radius Ratio & $R_{p,b}/R_*$ & Gaussian (0.02113, 0.00020) \\ 

         & Effective Temp. & $T_{\text{eff}}$ & Gaussian (6009, 64) \\
         & Metallicity & [$Fe/H$] & Gaussian (0.16, 0.04) \\
         & Stellar Radius & $R_{*}$ & Gaussian (1.482, 0.021) \\
         & Baseline Flux & $F_{0}$ & Gaussian (1.000, 0.001) \\
        \hline
    \end{tabular}
    \label{tab:exofast_priors}
\end{table*}

\begin{table}
    \centering
    \caption{Transit timing variations for the seven transits of the outer planet in the Kepler-103 system (Kepler-103c).}
    \begin{tabular}{cc}
        \hline
        Transit No. & Transit Timing Variation (mins)  \\
        \hline
        1 & $0.86^{+2.6}_{-2.4}$ \\
        2 & $-11{+4.3}_{-4.5}$ \\
        3 & $-23^{+3.2}_{-2.9}$ \\
        4 & $18\pm2.4$ \\
        5 & $10.2\pm2.4$ \\
        6 & $7.6^{+3.0}_{-2.9}$ \\
        7 & $-34\pm3.5$ \\
        \hline
    \end{tabular}
    \label{tab:ttv}
\end{table}

We performed transit fits to the photometric light-curve data of PH-2 and Kepler-103 using the publicly available MCMC software, {\tt EXOFAST v2} \citep{eastman2017}. The global model used in {\tt EXOFAST v2} includes spectral energy density and integrated isochrone models to constrain stellar parameters. In the case of both targets, we set the \emph{/torres} flag in order to utilize stellar mass and radius relations published in \cite{torres10}. The default limb darkening fit used by {\tt EXOFAST v2} is based on tables reported in \cite{claret11}. Relevant output parameters are listed in Tables \ref{tab:3663_output} and \ref{tab:103_output} respectively, the resulting transit fits are shown in Figure \ref{fig:kepler} for all three planets, and logistics of each fit are detailed below.

In the case of PH-2, we performed a fit using the five quarters of Kepler long-cadence light curves containing transits (Q4, Q7, Q10, Q13, Q16). We ran the MCMC fit for a maximum of 50,000 steps (\emph{maxsteps} = 50,000), recording every twentieth step value as part of the final posterior distribution (\emph{nthin} = 20). The default limb darkening fit failed, so we set the \emph{noclaret} flag to perform a solution ignoring the tables from \cite{claret11}. We performed a single-planet transit fit with open eccentricity, using the priors listed in Table \ref{tab:exofast_priors}. Priors on the two parameters related to the host star (effective temperature and metallicity) came from the analyses described in Section \ref{sec:stellar}, while the prior on eccentricity came from the RV fit, and was required to break a degeneracy between stellar density and eccentricity.

In the case of Kepler-103, we fit 18 quarters of long-cadence light-curves (Q0 - Q17), all containing transits. We ran the MCMC fit for a maximum of 5000 steps (i.e. \emph{maxsteps} = 5000), recording every hundredth step value as part of the final posterior distribution (i.e. \emph{nthin} = 100). We first performed a two-planet transit fit with both the planets fixed to circular orbits. In the first fit, we only applied priors on effective temperature, metallicity, and stellar radius, again provided by the analyses described in Section \ref{sec:stellar}. The prior on stellar radius broke a degeneracy between $R_{\ast}$ and the semi-amplitudes of the two planets in the system.  The circular fit produced white residuals for transits of Kepler-103b, however, a signal near ingress and egress of Kepler-103c in the residuals suggested transit timing variations (TTVs) were present. We therefore performed a second fit keeping Kepler-103b fixed to a circular obit, but allowing for TTVs and an eccentric orbit in the fit to transits of Kepler-103c and using the priors listed in Table \ref{tab:exofast_priors}. In the second fit that allowed for TTVs and an eccentric orbit on the second planet, we also placed priors on orbital periods, central transit times, baseline flux, and the star-to-planet radius ratio for Kepler-103b, based on successful parts of our first circular fit. The central transit time in each case is given in the {\it Kepler} format, meaning $T_{\text{c,BJD}} = T_{\text{c,KEP}} + 2454833$. TTV results from our second fit to transits of Kepler-103c are listed in Table \ref{tab:ttv}.

Relevant output parameters from these transit fits are given in Tables \ref{tab:3663_output} \& \ref{tab:103_output}, along with the results from the radial velocity analysis for each system, discussed in the next section.

%%%%%%%%%%%%%%%
% RV ANALYSIS %
%%%%%%%%%%%%%%%

\section{Radial velocity analysis} \label{sec:RV_analysis}

\subsection{Preliminary Investigation} \label{sec:RV_prelim}

\subsubsection{BGLS Periodograms} 

\begin{figure}
    \centering
        \includegraphics[width=9cm]{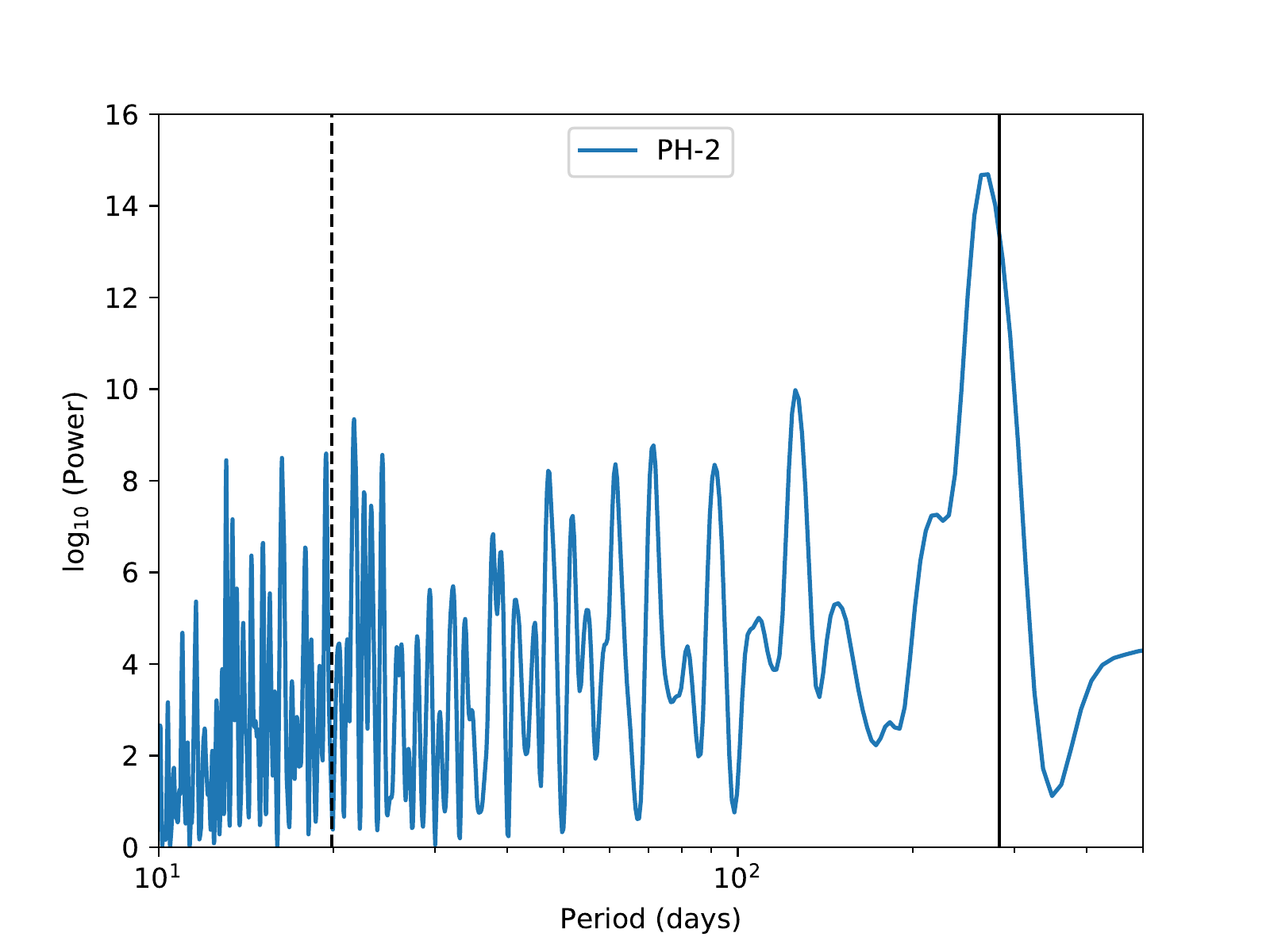}
        \includegraphics[width=9cm]{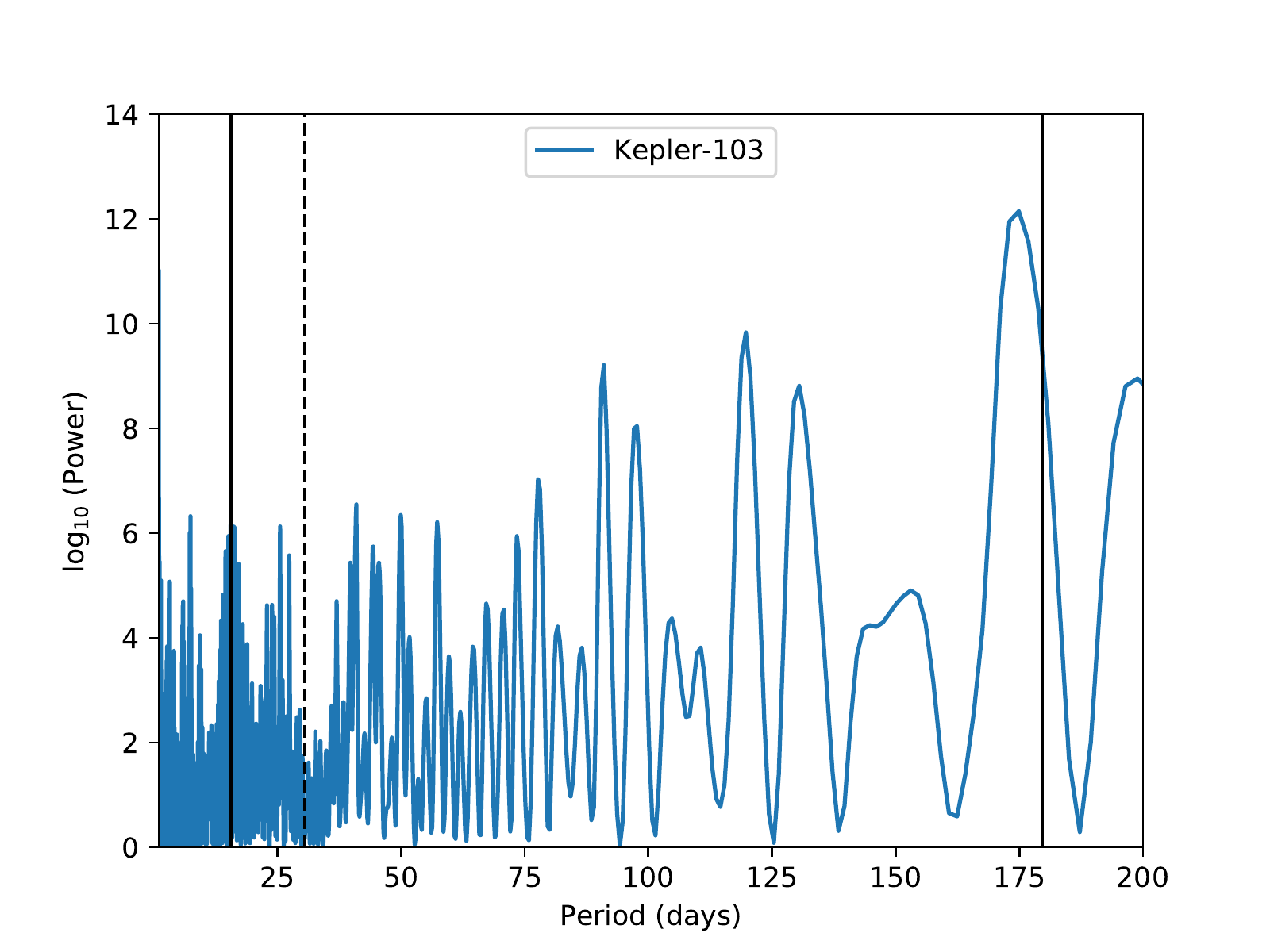}
    \caption{Bayesian generalised Lomb-Scargle periodograms of the RVs for PH-2 (top panel) and Kepler-103 (bottom panel). The solid lines indicate the expected periods of the known planets (PH-2b, Kepler-103b, and Kepler-103c).  There are clear signals at the periods of PH-2b ($P \sim 282.5$ days) and Kepler-103c ($P \sim 179.6$ days). There is also an indication of a signal at the period of Kepler-103b ($P \sim 15.7$ days). The dashed lines indicate expected rotation periods of the star.  There is some power at this period for PH-2, but little indication of a stellar rotation signal in the RVs of Kepler-103.}
    \label{fig:bgls}
\end{figure}

To check for the presence of planetary signals, and to gain an initial understanding of the activity of each star, we carried out a Bayesian generalised Lomb-Scargle (BGLS) periodogram analysis \citep{mortier15}. The results are presented in Figure \ref{fig:bgls}, and show period against the log of the power at each period. 

In each case, we looked for signs of periodicity matching the rotation period of the star (indicated by the dashed line). This would be a sign of stellar activity in the RVs, and may indicate that activity should be included in the RV analysis. Due to the noise levels of the data, and thus the clarity of the BGLS, it is unclear as to whether this is the case for either target. PH-2 shows some power at the expected stellar rotation, but there is little indication of a periodic stellar activity signal in the Kepler-103 RVs.  

The BGLS periodograms were also  used to confirm the existence of the planetary signals, the periods of which are indicated by the solid black lines. The top panel shows a very clear signal at the period of PH-2b ($P = 282.5$ days), while the bottom panel shows a very clear signal at the period of Kepler-103c ($P = 179.6$ days).  There is also an indication of a signal at the period of Kepler-103b ($P \sim 15.7$ days) in the bottom panel.  This gave confidence that we would be able to extract the radial velocity signatures of these planets. 

\subsubsection{Correlations} \label{sec:correlations}

\begin{table}
    \centering
    \caption{Pearson correlation coefficients for correlations between RV and the different activity indicators for both PH-2 and Kepler-103.}
    \begin{tabular}{lcccr}
        \hline
         ID & $r_{\text{BIS}}$ & $r_{\text{Contrast}}$ & $r_{\text{FWHM}}$ & $r_{\text{log} R'_{\text{HK}}}$ \\
         \hline
         PH-2 & 0.205 & 0.212 & 0.125 & 0.391 \\
         Kepler-103 & 0.165 & 0.117 & 0.153 & 0.019 \\
         \hline
    \end{tabular}
    \label{tab:pearson}
\end{table}

To further investigate the significance of the stellar activity, we looked at the correlations between the RVs and the activity indicators provided by the HARPS-N DRS (described in Section \ref{sec:obs_spec}). If any show a strong correlation, this can indicate that the RVs are strongly contaminated with signals relating to stellar activity.

The Pearson correlations coefficients, $r$, for correlations between RVs and four different activity indicators (BIS, contrast, FWHM, and $\log R'_{\text{HK}}$), are shown in Table \ref{tab:pearson}, for both PH-2 and Kepler-103. The $r$ values given for the $\log R'_{\text{HK}}$ were found using only non-zero values of this indicator. We consider any $r$ values greater than $\approx$ 0.5 as showing signs of significant correlation. This is not the case for any of the indicators for PH-2 or Kepler-103, a sign that the RV data does not support including stellar activity terms in either fit. We do note that the lack of correlations with these indices does not conclusively exclude stellar signals to be present in the data \citep[see e.g. ][]{cameron19}.

\subsection{Bayesian Analysis}\label{sec:methods_pyorbit}

To carry out the RV analysis, and to extract planetary parameters from the RV data, we used {\tt PyORBIT} \citep{malavolta16}.  {\tt PyORBIT} offers various options for the techniques used for each step of the analysis; here we used {\tt PyDE} \citep{storn97} for initial parameter determination, and {\tt emcee} \citep{foreman13} to do a Markov Chain Monte Carlo (MCMC) parameter estimation. We used uniform priors for the radial velocity semi-amplitude, $K$, and for the eccentricity, $e$, and used Gaussian priors for the orbital period, $P$, and for the mid-transit time, $T_{\rm cent}$. The priors for the orbital period and mid-transit time are taken from the photometry analysis discussed in Section \ref{sec:transit}, as is the inclination, which is taken to be fixed.

The RV model fit is defined by

\begin{equation}
\begin{aligned}
    RV(t) = K \cos(\omega + \nu(t)) +  K e \cos(\omega) + \gamma.
\end{aligned}
\label{eq:RV}
\end{equation}

Here, $\nu$ is the true anomaly, which is itself a function of $t,e,P$ and phase, $\omega$ is the argument of pericenter, and $\gamma$ is the RV offset. Following the recommendations in \citet{eastman13}, we used $\sqrt{e}\sin\omega$ and $\sqrt{e}\cos\omega$ as fitting parameters rather than $e$ and $\omega$. The period and semi-amplitude are explored in logarithmic space. A jitter term, $\sigma_{\text{jit}}$ is also added in quadrature to the model errors, to account for the white noise levels in the data.

% % % % % % % % % % % % % % % % % % % % % % % % % % % %
%       PH-2 Results
% % % % % % % % % % % % % % % % % % % % % % % % % % % %

\subsubsection{PH-2} \label{sec:results_ph2}

\begin{figure}
	\centering
	\includegraphics[width=9cm]{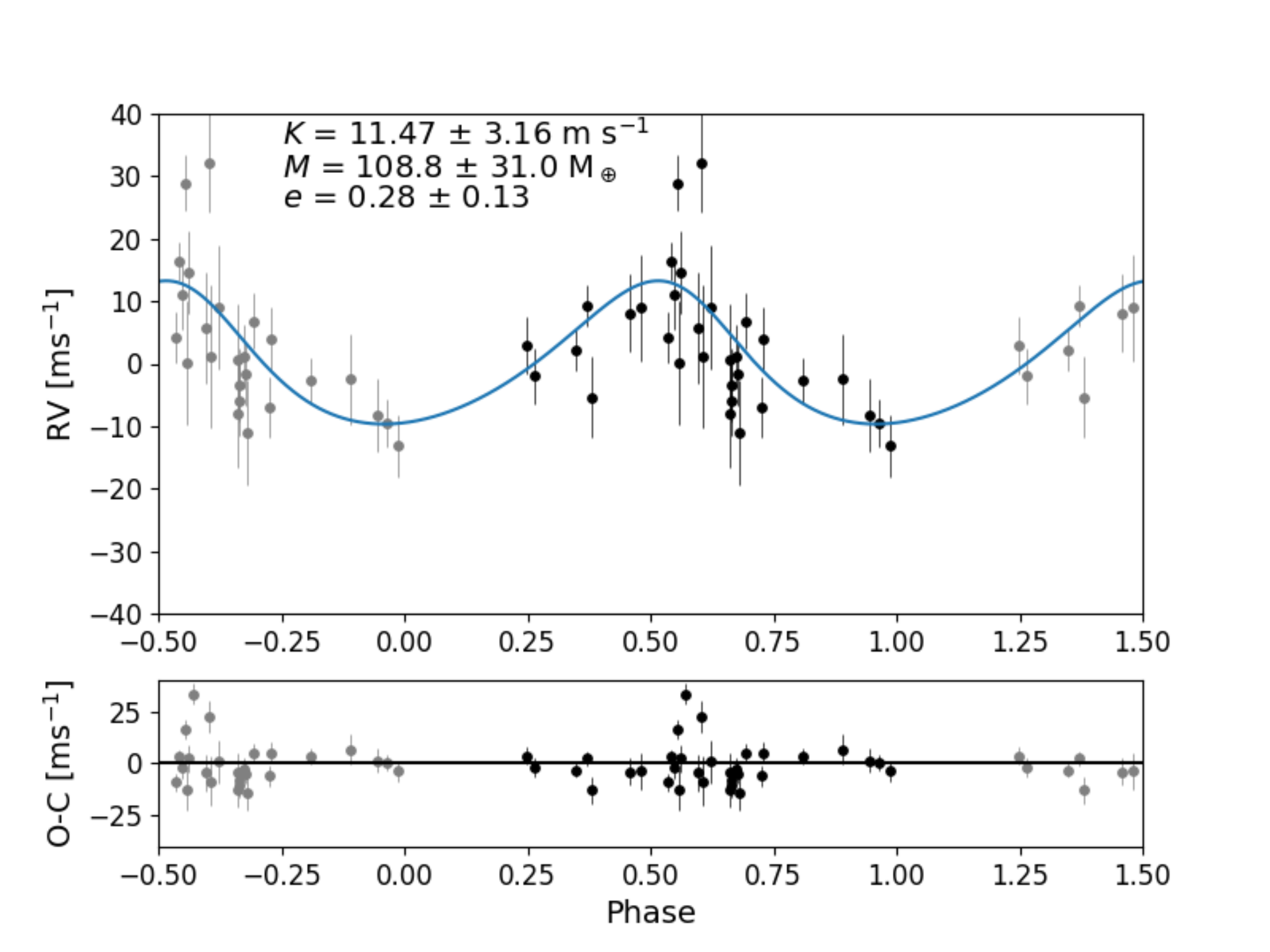}
	\caption{Phase-folded orbital solution (top) and RV residual (bottom) for PH-2b.}
	\label{fig:pyorbit_model_3663}
\end{figure} 

\citet{wang13} first reported the existence of a planet in the PH-2 system, discovered through the Planet Hunters project. Through false-probability analysis, they determined that the transit signal was caused by a giant planet at a 99.9 \% confidence level. However with only 4 RVs, they were unable to confirm this further. Our RVs and analysis here confirm the nature of the transit signal to be planetary.

The results of both the {\tt PyORBIT} RV model and the {\tt EXOFAST} transit model for PH-2 are shown in Table \ref{tab:3663_output}. The table shows the orbital period, $P$, and semi-major axis, $a$, in addition to the quantities derived from the RVs (radial velocity semi-amplitude, $K$, planet mass, $M_{\text{p}}$, eccentricity, $e$, and mean density, $\rho$). We also give the uncorrelated jitter and the RV offset. The posterior distributions of the derived quantities are shown in Figure \ref{fig:PH2_corners}. The expected lack of correlation between parameters is clearly visible, as is the convergence of the model to a final solution. 

Figure \ref{fig:pyorbit_model_3663} shows the orbital solution and residuals from the PH-2 RV analysis.  Our analysis recovers a RV semi-amplitude of $K = 11.47^{+3.01}_{-3.30}$ m s$^{-1}$, and an eccentricity of $e = 0.28^{+0.12}_{-0.13}$.   The errors on the final semi-amplitude (and thus mass) yield a better than 3$\sigma$ result.  Given the stellar mass of PH-2 and this RV semi-amplitude, we derive a mass for PH-2b of $M_{\text{p}} = 108.81^{+29.79}_{-32.29}$ M$_\oplus$.

The value of uncorrelated jitter in the final model fit is higher than one would expect for a slowly rotating, solar-type star \citep{cameron19}. This seems likely due to a combination of some level of stellar activity, and high levels of white noise caused by the irregular sampling of the planet's orbit. However, due to the quality of the data, we are unable to place any further constraints on the origin of this additional variability.

To test the viability of the model with such a high jitter, we also modelled the RV data without including a jitter term. The resulting RV semi-amplitude, $K = 13 \pm 1.3$ m s$^{-1}$, is consistent with the reported result. Whilst it is additionally more precise, we considered the Bayesian information criterion (BIC) of both models when determining which model to present. The BIC of the model including a jitter term is significantly lower than that which did not include jitter, with $BIC_{\text{jitter}} = 247.93$, compared to $BIC_{\text{no-jitter}} = 295.17$. As the difference between these two values is significantly larger than 10, it can be considered strong evidence that the model including a jitter term is favoured.

\begin{table}
	\centering
    \caption{Parameter Values for PH-2b from the RV and transit analyses}
	\begin{tabular}{l|l|l}
	    \hline
		Parameter & Description &  Value \\ 
		\hline
		{\it Transit analysis} & & \\
		\noalign{\smallskip}
		$P$ (days) & Orbital period & $282.52540^{+0.00010}_{-0.00011}$ \\
		\noalign{\smallskip}
		$T_{\text{cent}}$ (BJD) & Mid-transit time & $2455761.12272 \pm 0.00015$ \\
		\noalign{\smallskip}
		$R_p/R_*$ & Radius ratio & $0.09039^{+0.00051}_{-0.00055}$ \\
		\noalign{\smallskip}
		$R_p$ (R$_\oplus$)$^{a}$ & Planet radius & $9.49 \pm 0.16$ \\
		\noalign{\smallskip}
		$i$ (deg) & Inclination & $89.915^{+0.020}_{-0.022}$ \\ 
		\noalign{\smallskip}
		$a/R_*$ & Scaled semimajor axis & $185.76^{+3.89}_{-3.75}$ \\
		\noalign{\smallskip}	
		$a$ (au)$^{a}$ & Orbital semimajor axis & $0.824^{+0.019}_{-0.017}$ \\
		\noalign{\smallskip}
		\hline
		{\it RV analysis} & & \\
		\noalign{\smallskip}
		$K$ (m s$^{-1}$) & RV semi-amplitude & $11.47^{+3.01}_{-3.30}$ \\
		\noalign{\smallskip}
		$e$ & Eccentricity & $0.280^{+0.121}_{-0.133}$ \\
        \noalign{\smallskip}
		$\omega$ & Argument of pericenter &	$0.532^{+0.901}_{-0.796}$ \\
		\noalign{\smallskip}
		$M_p$ (M$_{\oplus}$) & Planet mass & $108.81^{+29.79}_{-32.29}$ \\
		\noalign{\smallskip}
		$\rho_p$ (g cm$^{-3}$) & Planet density  &  $0.70^{+0.20}_{-0.21}$ \\
		\noalign{\medskip}		
		$\sigma_{\rm jit}$ (m s$^{-1}$) & Uncorrelated jitter & $8.80^{+1.93}_{-1.62}$  \\
		\noalign{\smallskip}
		$\gamma$ (m s$^{-1}$) & RV offset & $-18733.56^{+2.45}_{-2.33}$ \\ 
		\noalign{\smallskip}
		\hline
	\end{tabular}
	\label{tab:3663_output}
\begin{flushleft}
{\bf Notes.}$^a$ Radius and semimajor axis are derived using our estimate for the stellar radius, $R_*=0.961^{+0.016}_{-0.015}$\,R$_{\odot}$, and the ratios $R_{\text{p}}/R_*$ and $a/R_*$ respectively. \\
\end{flushleft}
\end{table}

Combining the mass estimate with the \texttt{EXOFAST}-determined radius of $R_{\text{p}} = 9.49 \pm 0.16$ R$_{\oplus}$, leads to a density of $\rho_{\text{p}} = 0.70^{+0.20}_{-0.21}$ g cm$^{-3}$ = $0.13 \pm 0.04$ $\rho_{\oplus}$. This means that PH-2b has a mass and bulk density very similar to that of Saturn .  Using the median values, PH-2b has a mass of 1.14 Saturn masses and a density 1.03 times that of Saturn.  This strongly suggests that PH-2b is a Saturn-like gas giant.  

We also calculated the equilibrium temperature for PH-2b. We used the albedo of Saturn, $A_{\rm Sat} = 0.34$ \citep[page 61,][]{irwin03}, and a value of $f=1$, which is used for non-tidally locked planets that have uniform equilibrium temperatures across both hemispheres. This is a valid assumption, as the large orbital period of this planet means it is unlikely to be tidally locked. We also used the effective temperature, $T_{\text{eff}}$, from {\tt FASMA} (Table \ref{tab:stell_par}) and $a/R_{\ast}$ from the transit fit . Using  these parameters, we obtain a value of $T_{\text{eq}} = 251.87^{+3.772}_{-3.755}$ K.

PH-2 had been observed using RV instruments twice before, using the Keck-HIRES \citep{vogt94} and the SOPHIE \citep{perruchot08} spectrographs. The HIRES observations were taken in 2013, and are discussed in \citet{wang13}. 4 RV measurements were obtained, with formal internal uncertainties of $\sigma \approx $ 2 m/s. The RMS of this data is 14.0 m/s, indicating that the formal uncertainties were likely underestimated, and no constraints could be placed on properties of the orbiting companion of this star. However, based on statistical considerations of the properties of different types of companion, they did conclude that PH-2b should be a giant planet with a `very high likelihood'. The SOPHIE data consists of a further 4 RV points, discussed in \citet{santerne15}. The RMS of these points is 20.0 m/s, and they found a correlation (r=0.80) between the RV and the bisector span, which we do not recover in our analysis  (see Section \ref{sec:correlations}). Hence they concluded that they were also unable to confirm the planetary nature of the candidate. Thus, with the lower errors and RMS of HARPS-N, and by finding a \textgreater 3$\sigma$ mass value, this is the first work that is able to confidently say that the companion of PH-2 is indeed a planet.

In an attempt to improve the precision of our parameter values, we also included both the SOPHIE and HIRES data in the RV analysis. However, since the errors in both data sets were larger than those from HARPS-N, there was no significant improvement in the precision of the final parameters. As a result, we used only the HARPS-N data in our final analysis.

% % % % % % % % % % % % % % % % % % % % % % % % % % % % 
%       Kepler-103 Results
% % % % % % % % % % % % % % % % % % % % % % % % % % % % 

\subsubsection{Kepler-103}\label{sec:results_103} 

Two planets are known to be present in the Kepler-103 system, with current mass constraints of $M_{\text{b}}$ = 9.7 $\pm$ 6.8 M$_{\oplus}$ and $M_{\text{c}}$ = 36.1 $\pm$ 25.2 M$_{\oplus}$ \citep{marcy14}. Our goal was to use the HARPS-N RVs to reduce these uncertainties, and thus report the first precise mass measurements for these planets.

Table \ref{tab:103_output} gives the results of both the {\tt PyORBIT} RV model and the {\tt EXOFAST} transit model for Kepler-103.  We present the orbital period, semi-major axis and the quantities derived from the RVs (RV semi-amplitude, planet mass, eccentricity, and density) as well as the uncorrelated jitter and RV offset.  The posterior distributions of the derived quantities for Kepler-103b and Kepler-103c are shown in Figure \ref{fig:103_corners}.  Most of the parameters are well-constrained distributions, as one would expect for a converged solution. We find tails in both eccentricity distributions. However, the eccentricities for both Kepler-103b and Kepler-103c are also consistent with 0 at 2.45$\sigma$, which suggests that the eccentricity results are not significant \citep{lucy71}.  

From the RV fits, we find a 2$\sigma$ result for the RV semi-amplitude of Kepler-103b, and a 5$\sigma$ result for Kepler-103c. The derived planet masses are $M_{\text{p,b}} = 11.67^{+4.31}_{-4.73}$ M$_{\oplus}$ and $M_{\text{p,c}} = 58.47^{+11.17}_{-11.43}$ M$_{\oplus}$. These are consistent with, but more precise than, the results presented in \citet{marcy14}. 

Using these mass values and the planetary radii from the transit analysis, we obtain planetary densities of $\rho_{\text{b}} = 1.52^{+0.57}_{-0.61}$ g cm$^{-3}$ = $0.28 \pm 0.11 \rho_\oplus$ for Kepler-103b and $\rho_{\text{c}} = 1.98^{+0.44}_{-0.42}$ g cm$^{-3}$ = $0.36 \pm 0.08 \rho_\oplus$ for Kepler-103c. This suggests that Kepler-103b has a Neptune-like bulk density.  Kepler-103c, with a mass in between that of Neptune and Saturn, has no Solar System analogue.

For both planets, we again assumed a Saturn-like albedo. Using this, we estimate the equilibrium temperature, $T_{\text{eq}}$ of the planets to be $T_{\text{eq,b}} = 874.02^{+11.54}_{-11.41}$ K and $T_{\text{eq,c}} = 390.10^{+5.13}_{-5.10}$ K, for planets b and c respectively.

Kepler-103 has also been observed previously using the Keck-HIRES spectrograph \citet{vogt94}. \citet{marcy14} used 19 HiRES radial velocities to produce mass estimates for Kepler-103b and Kepler-103c. We did carry out an analysis that combined the HARPS-N RVs presented here, with the 19 HiRES RVs presented in \citet{marcy14}, but this did not improve the precision of our mass estimates. Hence, we report only on the results obtained using the HARPS-N RVs.

\begin{table*}
	\centering
    \caption{Parameter values from the Kepler-103 transit and RV analyses.}
	\begin{tabular}{l|l|l|l}
	    \hline
		Parameter & Description & Kepler-103b& Kepler-103c \\ 
		\hline
		{\it Transit analysis} & & & \\
		\noalign{\smallskip}
		$P$ (days) & Orbital period & $15.9653287^{+0.0000091}_{-0.0000092}$ & $179.60978^{+0.00019}_{-0.00020}$ \\
		\noalign{\smallskip}
		$T_{\text{cent}}$ (BJD) & Mid-transit time & $2455677.65243^{0.00028}_{-0.00028}$ & $2455667.15973^{0.00044}_{-0.00043}$ \\
		\noalign{\smallskip}
		$R_p/R_*$ & Radius ratio & $0.02140^{+0.00011}_{-0.00011}$ & $0.03351^{+0.0011}_{-0.00082}$ \\
		\noalign{\smallskip}
		$R_p$ (R$_\oplus$)$^{a}$ & Planet radius & $3.4857^{+0.0567}_{-0.0536}$ & $5.4539^{+0.1770}_{-0.1746}$  \\
		\noalign{\smallskip}
		$i$ (deg) & Inclination & $87.914^{+0.073}_{-0.072}$ & $89.704^{+0.12}_{-0.055}$ \\
		\noalign{\smallskip}
		$a/R_*$ & Scaled semimajor axis & $19.52^{+0.34}_{-0.34}$ & $98.0\pm1.7$ \\
		\noalign{\smallskip}
		$a$ (au)$^{a}$ & Orbital semimajor axis & $0.1330^{+0.0039}_{-0.0016}$  & $0.6679^{+0.0193}_{-0.0082}$ \\
		\noalign{\smallskip}
		\hline
		\noalign{\smallskip}
		{\it RV analysis} & & & \\
		\noalign{\smallskip}		
		
		\noalign{\smallskip}
		$K$ (m s$^{-1}$) & RV semi-amplitude & $2.75^{+1.05}_{-1.09}$  & $5.90^{+1.11}_{-1.13}$ \\
		\noalign{\smallskip}
		$e$ & Eccentricity & $0.171^{0.288}_{-0.124}$  & $0.103^{+0.092}_{-0.068}$\\
        \noalign{\smallskip}
		$\omega$ & Argument of pericenter &	$-0.815^{+2.301}_{-1.439}$  & $-0.272^{+1.467}_{-1.045}$ \\
		\noalign{\smallskip}
		$M_p$ (M$_{\oplus}$) & Planet mass & $11.67^{+4.31}_{-4.73}$ & $58.47^{+11.17}_{-11.43}$ \\
		\noalign{\smallskip}
		$\rho$ (g cm$^{-3}$) & Planet density & $1.52^{+0.57}_{-0.61}$ & $1.98^{+0.44}_{-0.42}$ \\
		\noalign{\smallskip}
		& & {\it Common parameters from RV analysis} \\
		\noalign{\smallskip}
		$\sigma_{\rm jit}$ (m s$^{-1}$) & Uncorrelated jitter & $1.93^{+1.03}_{-1.11}$ \\
		\noalign{\smallskip}
		$\gamma$ (m s$^{-1}$) & RV offset & $-28490.371^{+0.679}_{-0.678}$ \\   
		\noalign{\smallskip}
		\hline
	\end{tabular}
	\begin{flushleft}
	{\bf Notes.}$^a$ Radii and semimajor axes are derived using our estimate for the stellar radius, $R_*=1.492^{+0.024}_{-0.022}$\,R$_{\odot}$, and the ratios $R_p/R_*$ and $a/R_*$ respectively. \\
	\end{flushleft}
	\label{tab:103_output}
\end{table*}

\begin{figure}
	\centering
	\includegraphics[width=9cm]{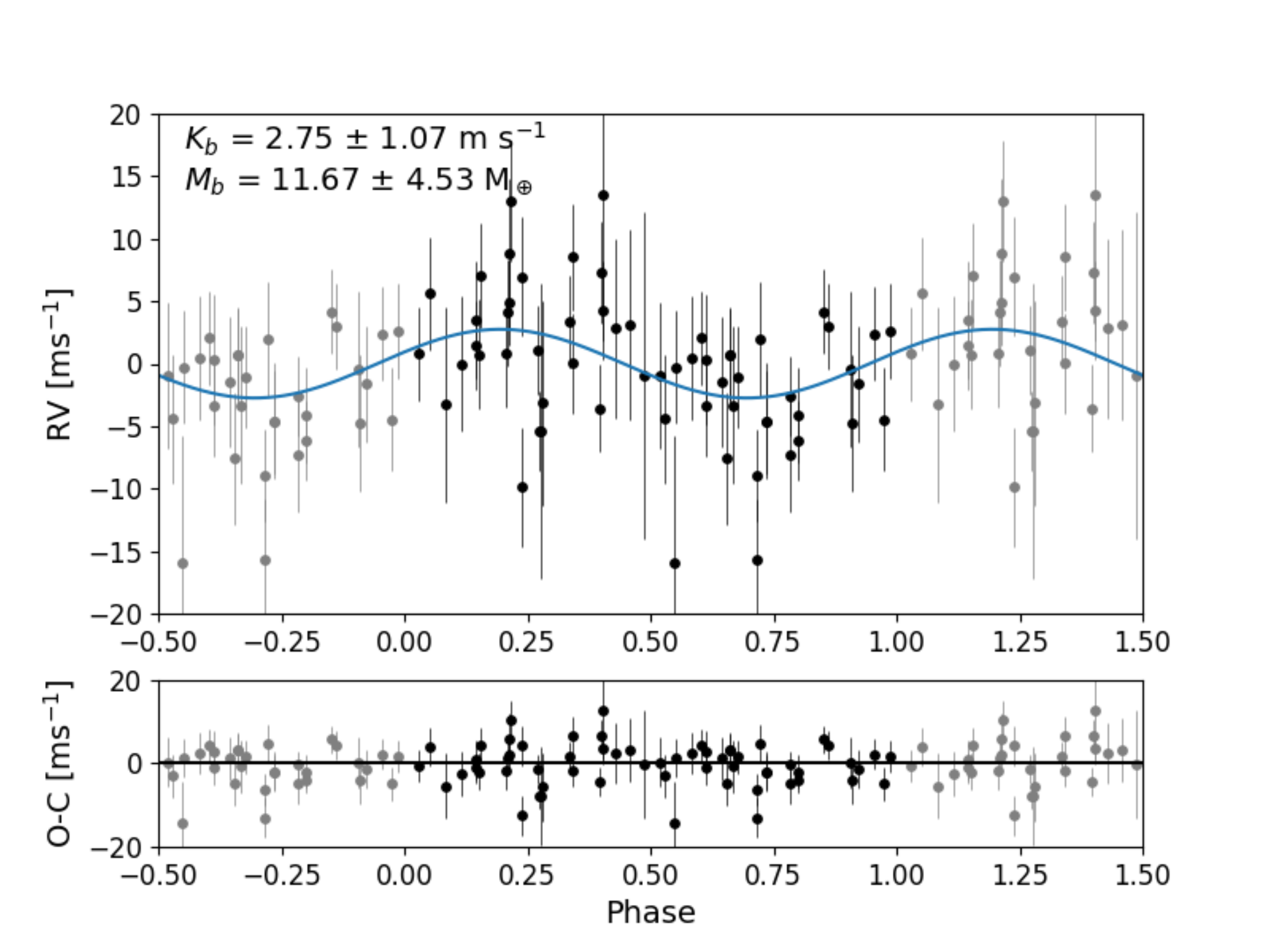}
	\includegraphics[width=9cm]{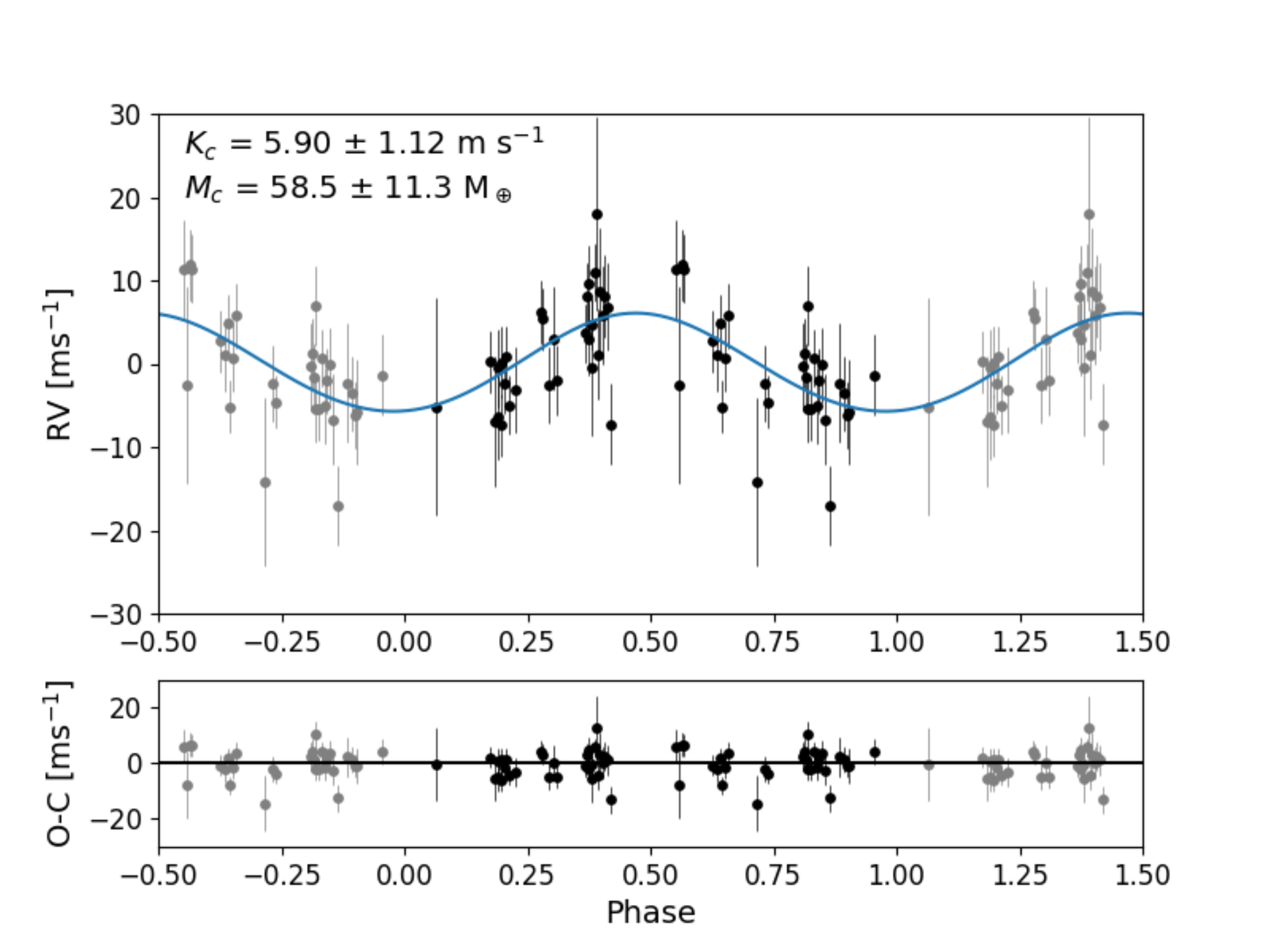}
	\caption{Orbital solutions and RV residuals for Kepler-103b (top) and Kepler-103c (bottom).  These are both phase folded on the period of the corresponding planet, and the RV contribution from the other planet has been removed.}
	\label{fig:pyorbit_model_103}
\end{figure} 

\subsubsection{Including stellar activity}
\label{sec:Incl_stell_acrt}

It is now well known that including stellar activity in the RV analysis can often improve the accuracy of the resulting mass estimates \citep[see][for an overview]{fischer16}. As discussed in Section \ref{sec:stell_act}, we carried out a GP analysis on the Kepler light-curves for PH-2 and for Kepler-103 and recovered rotation periods consistent with those determined through the ACF analysis.

We also carried out RV analyses in which we included stellar activity and assumed that the quasi-periodic kernel is the best choice to model the stellar activity induced RV variations.  We used the parameters determined in Section \ref{sec:stell_act} as priors for the stellar activity model.  

The results are consistent with those presented in Table \ref{tab:103_output}, but are much less precise.  This is either because, as discussed in Section \ref{sec:correlations}, the stellar activity is not strong enough to significantly influence the RVs in this dataset, or because the RV sampling is insufficient to constrain this activity. Consequently, since we cannot firmly comment on the activity levels of this star with this data, we present the results we obtained without including stellar activity in the RV analysis.  

%%%%%%%%%%%%%%
% DISCUSSION %
%%%%%%%%%%%%%%

\section{Discussion} \label{sec:disc}

\subsection{PH-2b} \label{sec:disc_ph2}

\begin{figure*}
	\centering
	\includegraphics[width=\linewidth]{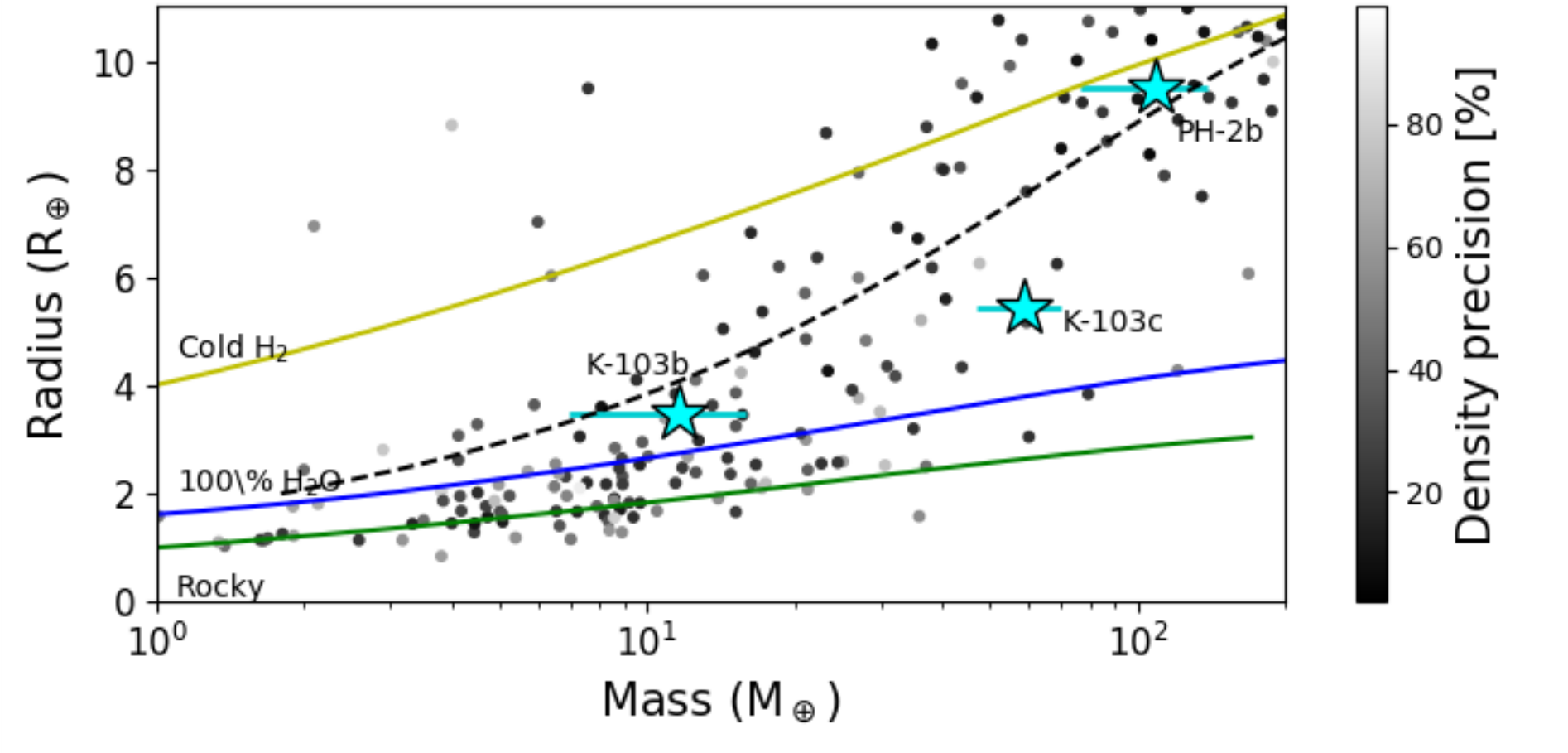}
	\caption{Planet radius versus planet mass for all known exoplanets with a radius smaller than 11 R$_\oplus$ and a mass smaller than 200 M$_\oplus$ (using \url{http://www.exoplanet.eu}, accessed 3 July 2019). The points are shaded with density precision where the darker points indicate the most precise measurements. PH-2b, Kepler-103b, and Kepler-103c are shown as cyan stars (note the errors for radius are smaller than the symbol). The solid lines represent planetary interior models for different compositions, top to bottom: H$_2$ assuming cold isentropic interior \citep{becker14,Zeng16}, 100\% H$_2$O assuming 1\,mbar surface pressure and 700\,K temperature, and Earthlike rocky \citep{Zeng19} The dashed line is a fit to a synthetic planet population for planets with $a>0.1$\,au \citep{mordasini12}.
	}
	\label{fig:mr}
\end{figure*}

\begin{figure*}
	\centering
	\includegraphics[width=\linewidth]{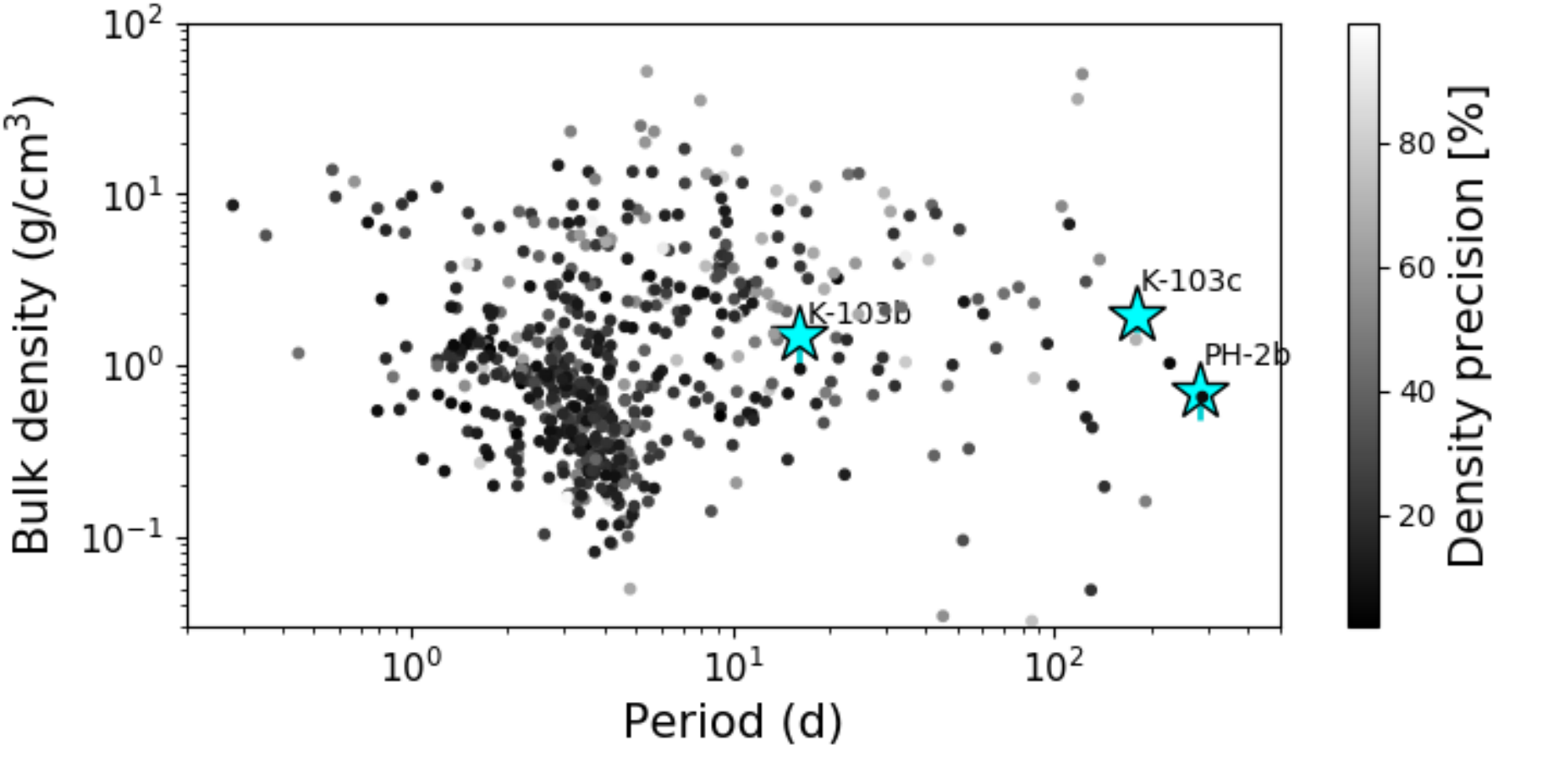}
	\caption{Planetary bulk density versus orbital period for all known exoplanets with measured densities (using \url{http://www.exoplanet.eu}, accessed 3 July 2019). The points are shaded with density precision where the darker points indicate the most precise measurements. PH-2b, Kepler-103b, and Kepler-103c are shown as cyan stars with errors. Note that most of these errors are smaller than the symbol)}
	\label{fig:per-dens}
\end{figure*}

Figure \ref{fig:mr} places PH-2b in the context of the larger sample of known exoplanets\footnote{Data from \url{http://www.exoplanet.eu}; accessed 3 July 2019}. We find that PH-2b fits in well with the other Saturn-like planets, even within the parameter errors. The planet's parameters nicely follow the mass-radius relation for longer-period giant planets from \citet{mordasini12}. We can also use its equilibrium temperature to infer that this planet is likely not highly irradiated. A large fraction of the well-studied population of hot Jupiters are known to have inflated radii compared to their longer-period counterparts \citep{demory11,sestovic18}. The possibility of this trend also being seen in the Saturn-like population is something that has been less investigated, due to the lack of planets that fall into this category. The NASA Exoplanet Archive \footnote{https://exoplanetarchive.ipac.caltech.edu/, accessed September 2019} lists 24 planets as having masses similar to Saturn (we considered masses in the range of 0.27 - 0.345 M$_{\text{J}}$ as Saturn-like). Of these, 20 are short-period planets, 19 of which have radii at least 10\% larger than Saturn. Of the remaining planets with approximately Saturn masses, only one is comparable to PH-2 in terms of period, Kepler-16b (discussed further below). It is also likely not inflated, possibly supporting the theory that the Saturn population displays the same behaviour as the Jupiters - i.e. the irradiated section of the population is significantly inflated when compared to the rest. Determining whether this is indeed the case is beyond the scope of this paper, as many more long-period Saturn-like planets are needed to properly distinguish between the two populations.

Figure \ref{fig:per-dens} demonstrates that our knowledge of long-period giant planets is sparse. PH-2b is one of only 4 planets with periods longer than 200 days, that have measured masses and radii. Of these, 2 have densities similar to PH-2b: Kepler-34b \citep{welsh12} and Kepler-16b \citep{doyle11}. Kepler-16b has an orbital period of 229 days, and a very well constrained density of $\rho = 1.031 \pm 0.015$ g cm$^{-3}$. It orbits an eclipsing binary system composed of a K5 main sequence star ($T_{\text{eff}} = 4450$K) and an M-type red dwarf, at semi-major axis of 0.7AU. The density of Kepler-34b is also well constrained: $\rho = 0.613^{+0.045}_{-0.041}$ g cm$^{-3}$, and it has a very similar orbital period to PH-2b, $P=289$\,days. Whilst the planet also orbits a binary star system, the binary components of Kepler-34 differ from those of Kepler-16 - two solar-type G stars with effective temperatures of $T_{\text{eff,A}} = 5913 \pm 130$K and $T_{\text{eff,B}} = 5867$K, respectively. Kepler-34b orbits at a semi-major axis of 1.0896 AU, significantly more distant from its host stars than PH-2b.
Both of these Saturn-like, long period planets are circumbinary. Consequently, PH-2b is the longest period Saturn-like planet that orbits a single star. If we choose to not use the period to inform our choices of comparison planets, we also find a strikingly similar planet with a period of 95 days. CoRoT-9b, reported by \citet{deeg10}, also orbits a single star and has a bulk density of 0.9 g cm$^{-3}$, statistically comparable to PH-2b. Furthermore, it is found to have an equilibrium temperature of 250-430K. The apparent resemblance of the two objects is another strong indication that the properties of these further out, less-irradiated planets may differ significantly from their closer-in analogues. 

The potential habitability of newly-discovered objects is often of great interest to the exoplanet community. One way of defining habitability is to consider the spectral type of the star (and thus temperature), and the distance of the planet from the star (semi-major axis, $a$). Using the methods of \citet{kopparapu13} (specifically Figure 7), within the errors on $M_{\ast}$ and $a$, this planet lies just on the edge of the habitable zone for this type of star.  There is still much uncertainty as to the most accurate method of defining habitable zones, and we refrain from reaching any concrete conclusions.

Furthermore, considering the habitability of a Saturn-like gas giant would be futile, but this is an interesting result when thinking about potential moons of planets such as this one. Some of the solar system moons \citep[e.g Enceladus -][]{waite17} are proving to be the most likely places for extra-terrestrial life to exist in our solar system, and so considering this possibility in extra-solar systems is also interesting. However, the potentially-habitable solar-system moons all reside beyond the habitable zone of the sun, and so the location of exoplanets should not be used to rule conclusively on whether they may have habitable moons.

Recently, \citet{guimaraes18} collated a list of the best candidate planets to host a detectable exomoon, using the full {\it Kepler} database. From 4417 objects that were analysed, they found 54 that were considered the best candidates for detecting the presence of an exomoon using the {\it Kepler} light-curve. PH-2 is among these candidates, identified as a likely candidate for detecting `icy' moons at a quarter of the maximum possible orbital distance from the planet. Our derived properties not only confirm this candidate as being a planet, but also support this idea of it being a likely exomoon host, due to the similarities of its properties to the solar system gas giants. Unfortunately, the TESS mission \citep{ricker15} will not be able to observe PH-2 (see \citet{christ19}, Figure 1), and so observations of possible exomoons will have to wait for the next generation of transit instruments, such as PLATO \citep{rauer2014}.

\subsection{Kepler-103} \label{sec:disc_k103}

Figure \ref{fig:mr} also shows the positions of the two planets in the Kepler-103 system in the mass-radius diagram. The planets agree with the general trend seen in mass-radius space. Kepler-103b coincides well with the assumed mass-radius relation from \citet{han2014}, who calculate masses using the mass-radius relations from \citet{weissmarcy14}, \citet{lissauer11} and \citet{mordasini12}. Kepler-103c, on the other hand, lies below the assumed mass-radius relation. However, when the likely intrinsic scatter in the data is considered \citep{wolfgang16}, combined with the fact that the mass-radius relations are not very well defined at this time, this discrepancy is probably not significant.

Similarly to PH-2b, Kepler-103c lies in a relatively sparse region of period-density parameter space, as evident in Figure \ref{fig:per-dens}. As of June 2019, there are 14 planets with periods longer than 100 days, with well-constrained densities (i.e. calculated using precise mass and radius measurements) \footnote{https://exoplanetarchive.ipac.caltech.edu/, accessed June 2019}. Of these planets, 3 have densities comparable to Kepler-103c: Kepler-1657b \citep{hebrard19}, Kepler-539b \citep{mancini16} and HD80606b \citep{naef01,moutou09}. These three planets have many parameter values in common: they all have masses similar, if not multiple times larger, than Jupiter, with Jupiter-like radii, and are at an orbital distance of approximately 0.5 AU. In contrast, Kepler-103c is sub-Jupiter in mass and radius and orbits at a distance of $a=0.66$ AU. This suggests it is likely a different kind of object to the other long-period transiting planets characterised so far. 
Moreover, the properties of these systems can allow us to speculate on possible formation processes for Kepler-103b and c. The high eccentricity ($e=0.5\pm0.03$) of Kepler-1657b has been attributed to planet-planet interactions, after disk dissipation \citep{hebrard19}. HD80606b is likely extremely eccentric ($e=0.934\pm0.603$), but as it is in a binary star system this has been attributed to interactions with the outer stellar companion. Both of these systems are examples of formation mechanisms that have left strong detectable traces on the resulting planets, which have likely undergone dynamical orbital evolution. On the other hand, the eccentricities of both Kepler-103b and Kepler-103c are consistent with zero at a $\sim 1\sigma$ level. This suggests a very different evolution history, one that may not involve strong dynamical interactions. Furthermore, these systems are all thought to be of similar ages ($\sim 3.5$ Gyr, albeit with extremely large error bars). As a result, it is unlikely that the discrepancy between the properties of these systems can be attributed to each being at different stages in their evolution.

Due to the increased stellar mass and thus effective temperature of Kepler-103, as well as the smaller semi-major axes of the planets in the system, neither Kepler-103b or Kepler-103c lies in the predicted habitable zone of their host star, again according to the results of \citet{kopparapu13}.

In Section \ref{sec:transit}, we presented the TTVs from the Kepler-103 system in Table \ref{tab:ttv}. It was also was previously identified by \citet{holczer16} as a system with `significant long-term TTVs', with the amplitude of the variation found to be $\sim$ 15 minutes. It is thought that the majority of periodic TTVs are indicative of dynamical interactions with another planet in the system. In this case the possible third planet is likely to be non-transiting, further implied by the discrepancy in inclination (1.774 deg) between the two transiting planets Kepler-103b and Kepler-103c. However, as we only have 7 TTV measurements to analyse, it is challenging to obtain a good estimate for the period of this potential third planet, which would be necessary to include it in a further RV fit. We also saw no clear sign of another RV signal in the residuals from the RV fit. As a result, we leave this analysis for a future investigation of this system.

\section{Conclusions} \label{sec:concl}

In this work, we have confirmed the existence of a planet in the PH-2 system, and present a 3.5$\sigma$ mass estimate. We find that PH-2b has a mass of $M_{\text{b}} = 109^{+30}_{-32}$\,M$_{\oplus}$, and a radius of $R_{\text{b}} = 9.49 \pm 0.16 $\,R$_{\oplus}$. This suggests that PH-2b has a similar bulk density to Saturn, with $\rho_{\text{b}} = 1.02^{+0.29}_{-0.31}$\,$\rho_{\text{S}}$. This is also the first confirmed planet with a Saturn-like mass, that has a period of longer than 200 days, and that does not orbit a binary star system. 

We also recover the first precise mass estimates for the two known planets in the Kepler-103 system.  Kepler-103b is found to have a mass of $M_{\text{p,b}} = 11.7 \pm 4.5$ M$_\oplus$ and a density of $\rho_{\text{p,b}} = 1.52^{+0.57}_{-0.61}$ g cm$^{-3}$, while for Kepler-103c we find $M_{\text{p,c}} = 58.5 \pm 11.3$ M$_\oplus$ and $\rho_{\text{p,c}} = 1.98^{+0.44}_{-0.42}$ g cm$^{-3}$.  This suggests that Kepler-103b has a bulk density similar to Neptune, while Kepler-103c, with a period of $P_{\text{c}} = 179$ days, is one of the densest known long-period exoplanets. It also has no solar-system analogue in terms of density, and so could be a very interesting object to focus further observations on.

These results increase the sample of long-period ($P > 100$ days), intermediate-mass planets with well-constrained mass and radius estimates.  This is key in constraining the mass-radius relation across the full range of exoplanet masses and radii, and for gaining a better understanding of the processes involved with planet formation and evolution.

\section*{Acknowledgements}

We would like to thank the anonymous referee for a thorough report that improved the quality of this paper.

The HARPS-N project has been funded by the Prodex Program of the Swiss Space Office (SSO), the Harvard University Origins of Life Initiative (HUOLI), the Scottish Universities Physics Alliance (SUPA), the University of Geneva, the Smithsonian Astrophysical Observatory (SAO), and the Italian National Astrophysical Institute (INAF), the University of St Andrews, Queen's University Belfast, and the University of Edinburgh.

This research has made use of the SIMBAD database, operated at CDS, Strasbourg, France, and NASA's Astrophysics Data System.

Based on observations made with the Italian {\it Telescopio Nazionale
Galileo} (TNG) operated by the {\it Fundaci\'on Galileo Galilei} (FGG) of the {\it Istituto Nazionale di Astrofisica} (INAF) at the
{\it  Observatorio del Roque de los Muchachos} (La Palma, Canary Islands, Spain).

This paper includes data collected by the {\it Kepler}\ mission. Funding for the {\it Kepler}\ mission is provided by the NASA Science Mission directorate. Some of the data presented in this paper were obtained from the Mikulski Archive for Space Telescopes (MAST). STScI is operated by the Association of Universities for Research in Astronomy, Inc., under NASA contract NAS5-26555. Support for MAST for non-HST data is provided by the NASA Office of Space Science via grant NNX13AC07G and by other grants and contracts.

This work has made use of data from the European Space Agency (ESA) mission Gaia (https://www.cosmos.esa.int/gaia), processed by the Gaia Data Processing and Analysis Consortium (DPAC, https://www.cosmos.esa.int/web/gaia/dpac/consortium). Funding for the DPAC has been provided by national institutions, in particular the institutions participating in the Gaia Multilateral Agreement.

AM acknowledges support from Senior Kavli Institute Fellowships at the University of Cambridge.
ACC acknowledges support from the Science \&\ Technology Facilities Council (STFC) consolidated grant number ST/R000824/1.
AV's and RDH's work was performed under contract with the California Institute of Technology/Jet Propulsion Laboratory funded by NASA through the Sagan Fellowship Program executed by the NASA Exoplanet Science Institute.
LM acknowledges support from  PLATO ASI-INAF agreement n.2015-019-R.1-2018
This publication was made possible through the support of a grant from the John Templeton Foundation. The opinions expressed in this publication are those of the authors and do not necessarily reflect the views of the John Templeton Foundation. 
This material is partly based upon work supported by the National Aeronautics and Space Administration under grants No. NNX15AC90G and NNX17AB59G issued through the Exoplanets Research Program.
Some of this work has been carried out in the frame of the National Centre for Competence in Research `PlanetS' supported by the Swiss National Science Foundation (SNSF).

%%%%%%%%%%%%%%%%%%%%%%%%%%%%%%%%%%%%%%%%%%%%%%%%%%



\bibliographystyle{mnras}
\bibliography{project.bib} % if your bibtex file is called example.bib

%%%%%%%%%%%%%%%%%%%%%%%%%%%%%%%%%%%%%%%%%%%%%%%%%%

%%%%%%%%%%%%%%%%% APPENDICES %%%%%%%%%%%%%%%%%%%%%
\appendix

\section{Full Kepler light-curves}

\begin{figure*}
    \centering
    \includegraphics[width=0.75\textwidth]{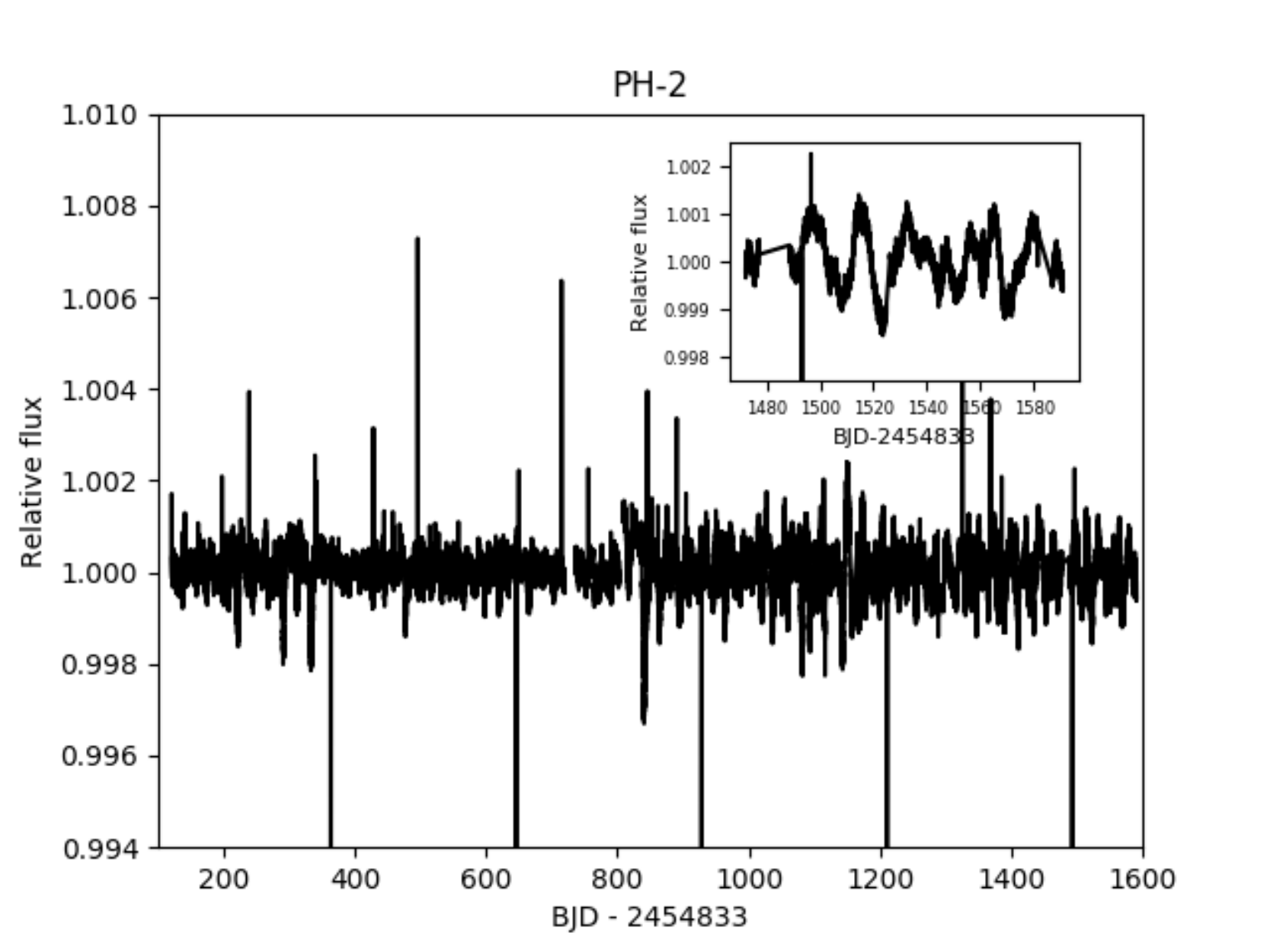}
    \caption{Full {\it Kepler} light-curve, including all quarters in which PH-2 was observed. Inset shows enhancement of final section of light-curve, to make any variability in the signal easier to identify visually.}
    \label{fig:PH2_full_lightcurve}
\end{figure*}

\begin{figure*}
	\centering
	\includegraphics[width=0.75\textwidth]{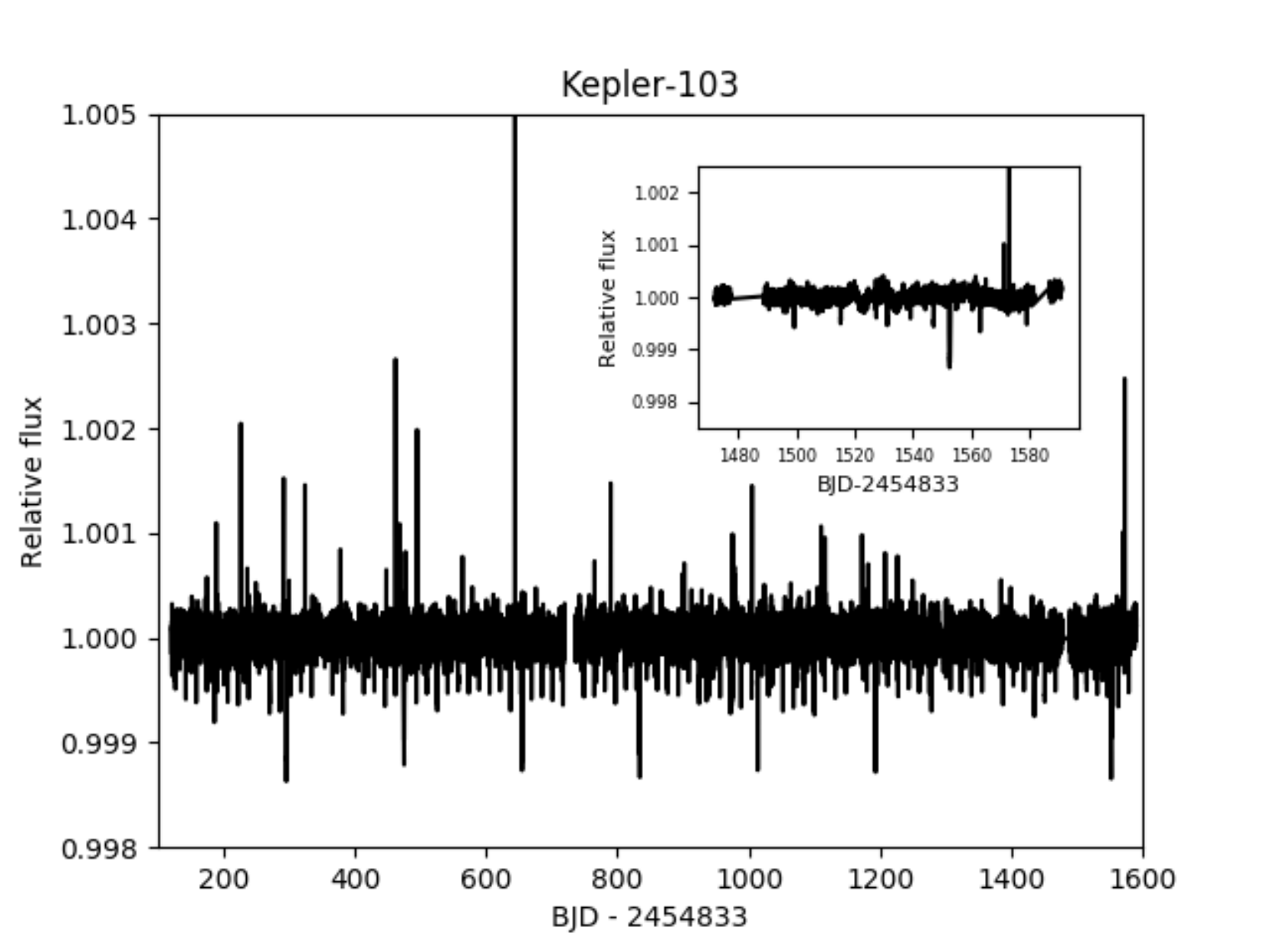}
    \caption{Full {\it Kepler} light-curve, including all quarters in which Kepler-103 was observed. Inset shows enhancement of final section of light-curve, to make any variability in the signal easier to identify visually.}
	\label{fig:K103_full_lightcurve}
\end{figure*}

\section{RV Data for PH-2}

\begin{table*}
    \centering
    \caption{Radial Velocity measurements for PH-2, with associated 1$\sigma$ errors. Also given are the activity indicators; FWHM, contrast and bisector span.}
	\begin{tabular}{lcccccr}
	    \hline
		Time & RV & Error & FWHM & Contrast & Bisector Span & log $R'_\text{HK}$ \\
		(BJD-2400000) & (km/s) & (km/s) & (km/s) & (\%) & (km/s) &  \\
        \hline
        56565.459650 & -18.71710 & 0.00305 & 7.02821 & 45.042 & -0.01724 & -4.7853	\\
        56567.458097 & -18.72263 & 0.00539 & 7.03170 & 44.857 & -0.01875 & -4.7275	\\
        56569.517547 & -18.70465 & 0.00448 & 7.03754 & 44.920 & -0.01694 & -4.8869	\\
        56571.481316 & -18.71900 & 0.00657 & 7.06412 & 44.631 & -0.00665 & -4.6800	\\
        56581.411323 & -18.72774 & 0.00888 & 7.02278 & 44.066 & -0.01064 & -4.6768	\\
        56583.363421 & -18.70145 & 0.00787 & 7.02456 & 43.151 & 0.04252	 & -4.7448	\\
        56604.361465 & -18.73521 & 0.00497 & 7.02704 & 44.949 & -0.01146 & -4.7726	\\
        56608.359816 & -18.72675 & 0.00447 & 7.02449 & 44.653 & -0.02649 & -4.7671	\\
        56765.708349 & -18.73062 & 0.00451 & 6.95605 & 45.282 & -0.02636 & -4.8251	\\
        56769.720051 & -18.73557 & 0.00454 & 6.93700 & 45.359 & -0.01887 & -4.8919	\\
        56793.619380 & -18.73142 & 0.00336 & 6.97480 & 45.833 & -0.01683 & 0.0000	\\
        56799.707356 & -18.72427 & 0.00332 & 6.93672 & 45.522 & -0.01988 & -4.8886	\\
        56802.652161 & -18.73891 & 0.00656 & 6.95788 & 45.418 & -0.03348 & -4.8140	\\
        56824.570925 & -18.72548 & 0.00626 & 6.96243 & 48.380 & -0.03417 & 0.0000	\\
        56830.565717 & -18.72464 & 0.00853 & 6.94560 & 45.295 & -0.01587 & -4.9963	\\
        56846.571444 & -18.72934 & 0.00412 & 6.93462 & 45.517 & -0.01412 & -4.8536	\\
        56852.520256 & -18.73340 & 0.00981 & 6.90791 & 44.359 & -0.01310 & -4.8677	\\
        56866.657993 & -18.73243 & 0.01149 & 6.96797 & 45.239 & -0.01257 & -5.2592	\\
        56885.561643 & -18.73247 & 0.00501 & 6.92626 & 45.358 & -0.03742 & -4.8490	\\
        56887.487825 & -18.74467 & 0.00844 & 6.92345 & 45.431 & -0.02491 & -4.7925	\\
        56900.543502 & -18.74056 & 0.00482 & 6.98739 & 45.105 & -0.03028 & 0.0000	\\
        56923.454772 & -18.73635 & 0.00372 & 6.94069 & 45.508 & -0.03071 & -4.8479	\\
        56967.395942 & -18.74305 & 0.00378 & 6.93866 & 45.007 & -0.01849 & -4.8905	\\
        57153.627419 & -18.72450 & 0.00996 & 6.96027 & 44.989 & -0.04658 & -4.7587	\\
        57164.712545 & -18.74168 & 0.00840 & 6.96791 & 46.070 & 0.00501	 & 0.0000	\\
        57164.723159 & -18.73298 & 0.00892 & 6.98420 & 45.234 & -0.03970 & 0.0000	\\
        57165.700073 & -18.73696 & 0.00577 & 6.95299 & 45.595 & -0.02610 & 0.0000	\\
        57165.710733 & -18.73948 & 0.00570 & 6.92862 & 45.551 & -0.01027 & 0.0000	\\
        57183.698281 & -18.72963 & 0.00506 & 6.91683 & 45.609 & 0.00562	 & -4.8693	\\
        57229.606372 & -18.73604 & 0.00724 & 6.90945 & 45.566 & -0.03378 & -5.0336	\\
        57256.473584 & -18.74672 & 0.00491 & 6.93905 & 45.579 & -0.03114 & -4.8579	\\
        57527.638124 & -18.74182 & 0.00588 & 6.91374 & 45.114 & -0.01815 & -4.8576	\\
	\end{tabular}
	\label{tab:rv_3663}
\end{table*}

\section{RV Data for Kepler-103}

\begin{table*}
	\centering
	\caption{Radial Velocity measurements for Kepler-103, with associated 1$\sigma$ errors. Also given are the activity indicators; FWHM, contrast and bisector span.}
	\begin{tabular}{lcccccr}
	    \hline
		Time & RV & Error & FWHM & Contrast & Bisector Span & log $R'_\text{HK}$ \\
		(BJD-2400000) & (km/s) & (km/s) & (km/s) & (\%) & (km/s) &  \\
		\hline
        56830.522954 & -28.49935 & 0.00527 & 8.04057 & 41.131 & 0.04491	& 0.0000 \\	
        56831.455893 & -28.50049 & 0.00371 & 8.03551 & 41.179 & 0.01920	& 0.0000 \\	
        56845.651341 & -28.48650 & 0.00374 & 8.05067 & 41.116 & 0.01817	& 0.0000 \\	
        56846.549929 & -28.48771 & 0.00372 & 8.04796 & 41.013 & 0.02187	& 0.0000 \\	
        56848.539131 & -28.49526 & 0.00464 & 8.04346 & 40.854 & 0.02140	& 0.0000 \\	
        56850.502033 & -28.48802 & 0.00620 & 8.05904 & 40.276 & 0.03447	& 0.0000 \\	
        56851.543536 & -28.49188 & 0.00416 & 8.01287 & 40.811 & 0.02295	& 0.0000 \\	
        56862.546688 & -28.48491 & 0.00388 & 8.04514 & 41.075 & 0.02585	& 0.0000 \\	
        56863.509442 & -28.48349 & 0.00458 & 8.02891 & 41.071 & 0.03017	& 0.0000 \\	
        56864.515598 & -28.48797 & 0.00319 & 8.05476 & 41.069 & 0.01835	& 0.0000 \\	
        56865.557066 & -28.48103 & 0.00340 & 8.03082 & 41.134 & 0.02961	& 0.0000 \\	
        56866.505409 & -28.48989 & 0.00546 & 8.03230 & 41.110 & 0.01916	& 0.0000 \\	
        57186.538255 & -28.48996 & 0.00373 & 8.04143 & 41.097 & 0.01879	& -5.1056 \\	
        57188.591209 & -28.49526 & 0.00782 & 8.05832 & 40.919 & -0.00053 & -4.9959 \\	
        57189.582383 & -28.48829 & 0.00468 & 8.01046 & 41.084 & 0.00250	& -5.2806 \\	
        57190.595039 & -28.48750 & 0.00430 & 8.04686 & 41.050 & 0.01530	& -5.0016 \\	
        57191.596062 & -28.49029 & 0.00358 & 8.03995 & 41.159 & 0.02140	& -5.1231 \\	
        57192.592662 & -28.48781 & 0.00366 & 8.04105 & 41.142 & 0.02154	& -5.1253 \\	
        57193.595165 & -28.49452 & 0.00347 & 8.01704 & 41.163 & 0.02139	& -5.0801 \\	
        57195.699998 & -28.49494 & 0.00517 & 8.03891 & 41.005 & 0.02516	& -5.1028 \\	
        57221.484029 & -28.48414 & 0.00355 & 8.02363 & 41.136 & 0.01315	& -5.1032 \\	
        57222.458119 & -28.48462 & 0.00438 & 8.01884 & 41.115 & 0.03077	& -5.0597 \\	
        57223.642615 & -28.48850 & 0.00818 & 8.04911 & 40.999 & -0.00467 & -4.9447\\	
        57225.612273 & -28.47164 & 0.01167 & 8.08135 & 40.815 & 0.03530	& -4.9240 \\	
        57226.498513 & -28.48195 & 0.00762 & 8.04607 & 40.936 & 0.02369	& -5.0481 \\	
        57227.474011 & -28.48585 & 0.00589 & 8.03837 & 40.968 & 0.00990	& -4.9999 \\	
        57228.500203 & -28.48439 & 0.00495 & 8.06543 & 40.942 & 0.03284	& -4.9124 \\	
        57229.498107 & -28.48617 & 0.00521 & 8.04793 & 40.983 & -0.00042 & -5.2145 \\	
        57230.616555 & -28.50034 & 0.00481 & 8.04698 & 41.118 & 0.02218	& -4.9834 \\	
        57254.491769 & -28.47638 & 0.00599 & 8.02801 & 41.193 & 0.01926	& -5.2237 \\	
        57255.552280 & -28.49063 & 0.01182 & 8.03905 & 40.722 & 0.00095	& -4.9006 \\	
        57256.569353 & -28.47688 & 0.00422 & 8.03408 & 41.144 & 0.03198	& -4.9985 \\	
        57257.504598 & -28.47821 & 0.00406 & 8.04376 & 41.128 & 0.03318	& -5.1546 \\	
        57267.529454 & -28.48629 & 0.00375 & 8.02966 & 41.050 & 0.01504	& -5.0856 \\	
        57269.488006 & -28.48670 & 0.00439 & 8.03324 & 41.109 & 0.02957	& -5.2216 \\	
        57270.475661 & -28.48274 & 0.00341 & 8.05041 & 41.101 & 0.02451	& -5.0874 \\	
        57271.478581 & -28.49319 & 0.00313 & 8.03565 & 41.115 & 0.01132	& -5.1177 \\	
        57272.517761 & -28.48795 & 0.00401 & 8.04563 & 41.088 & 0.00902	& -5.0489 \\	
        57273.494210 & -28.48389 & 0.00391 & 8.01073 & 41.152 & 0.01164	& -5.0586 \\	
        57301.455530 & -28.48644 & 0.00417 & 7.99831 & 41.120 & 0.01274	& -5.0875 \\	
        57302.456156 & -28.48065 & 0.00479 & 8.03336 & 41.095 & 0.03196	& -5.0952 \\	
        58361.450610 & -28.50616 & 0.01010 & 8.01820 & 40.852 & -0.01878 & 0.0000 \\	
        58364.470638 & -28.49544 & 0.00462 & 8.02952 & 41.060 & 0.02362	& -5.0950 \\	
        58365.473987 & -28.49722 & 0.00315 & 8.04112 & 41.128 & 0.01816	& -5.1476 \\	
        58378.462475 & -28.49312 & 0.00522 & 8.04947 & 41.064 & 0.01771	& -5.2483 \\	
        58379.483929 & -28.49472 & 0.00406 & 8.01745 & 41.054 & 0.03016	& -5.0245 \\	
        58380.425537 & -28.49843 & 0.00409 & 8.03422 & 41.096 & 0.01573	& -5.0580 \\	
        58381.449200 & -28.49799 & 0.00384 & 8.02395 & 41.090 & 0.02372	& -5.0569 \\	
        58382.431951 & -28.49110 & 0.00342 & 8.02684 & 41.062 & 0.02315	& -5.0229 \\	
        58383.432779 & -28.49581 & 0.00463 & 8.04518 & 40.902 & 0.02256	& -5.0624 \\	
        58384.432912 & -28.49176 & 0.00383 & 8.02738 & 41.007 & 0.00215	& -5.0250 \\	
        58385.449051 & -28.48885 & 0.00454 & 8.00777 & 40.767 & 0.02350	& -5.0092 \\	
        58386.511959 & -28.49465 & 0.00539 & 8.00206 & 40.318 & 0.01328	& -4.9821 \\	
        58388.481621 & -28.50472 & 0.00476 & 7.99761 & 40.893 & 0.01747	& -5.0013 \\	
        58391.497742 & -28.49238 & 0.00714 & 8.03932 & 40.814 & 0.03380	& -4.9110 \\	
        58393.444127 & -28.49567 & 0.00450 & 8.03522 & 41.206 & 0.02414	& -5.0629 \\	
        58394.427834 & -28.49889 & 0.00404 & 8.03380 & 41.195 & 0.01975	& -5.0829 \\	
        58395.334495 & -28.49886 & 0.00626 & 8.03863 & 40.981 & 0.02270	& -5.1372 \\	
        58404.414404 & -28.48907 & 0.00479 & 8.02798 & 41.161 & 0.01890	& -5.1170 \\	
        58424.332625 & -28.49622 & 0.01305 & 8.02904 & 40.997 & 0.03783	& -5.1329 \\
		\hline
	\end{tabular}
	\label{tab:rv_108}
\end{table*}

\section{Corner plots for the RV analyses}

\begin{figure*}
	\centering
	\includegraphics[width=0.95\textwidth]{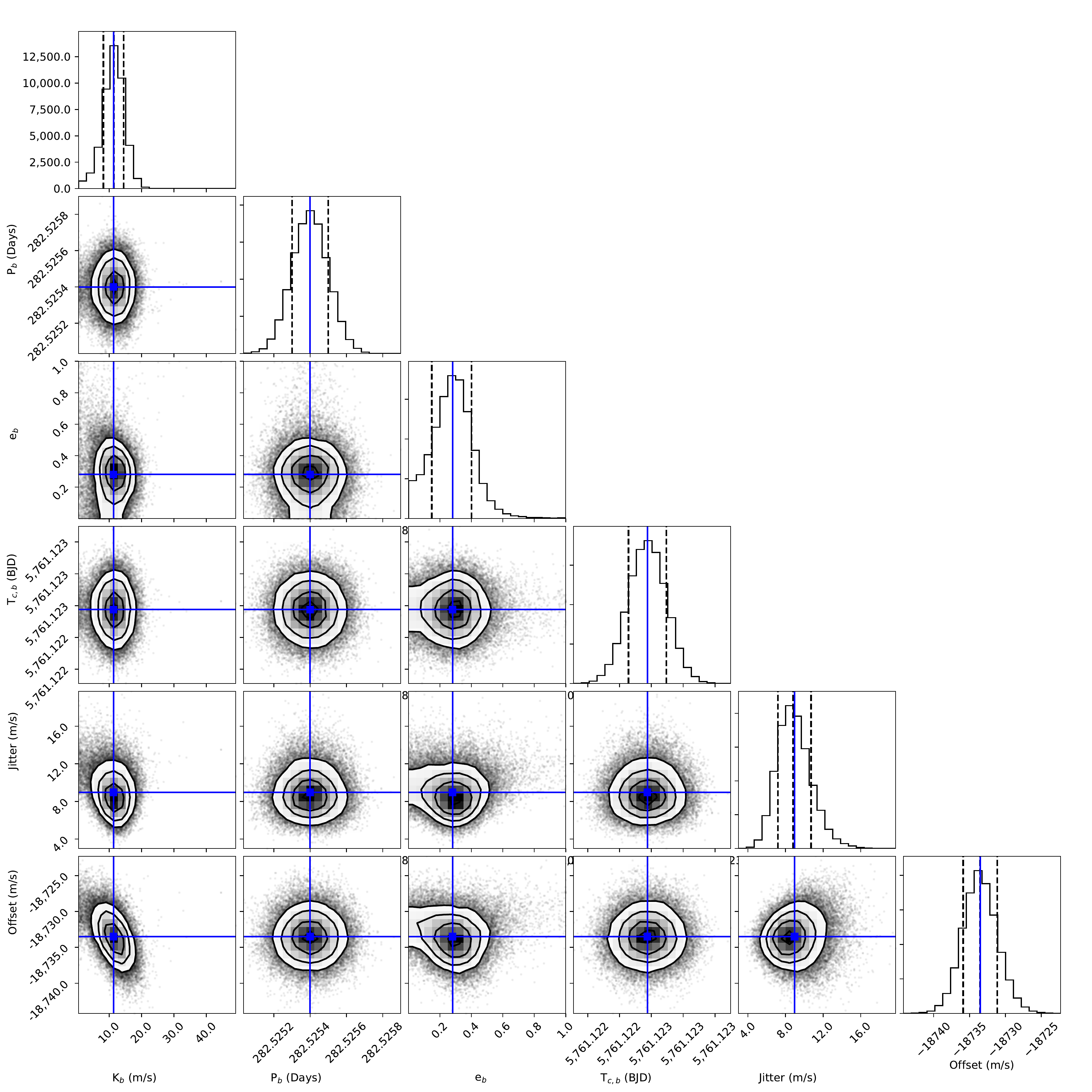}
    \caption{Posterior distributions for the internal variables determined from the PH-2 RV analysis. The contours are at $1$, $2$, and $3 \sigma$. Median values for each parameter denoted by the solid blue lines are those given in Table \ref{tab:3663_output}. Made using {\tt corner} \citep{corner}.}
	\label{fig:PH2_corners}
\end{figure*}

\begin{figure*}
	\centering
	\includegraphics[width=0.95\textwidth]{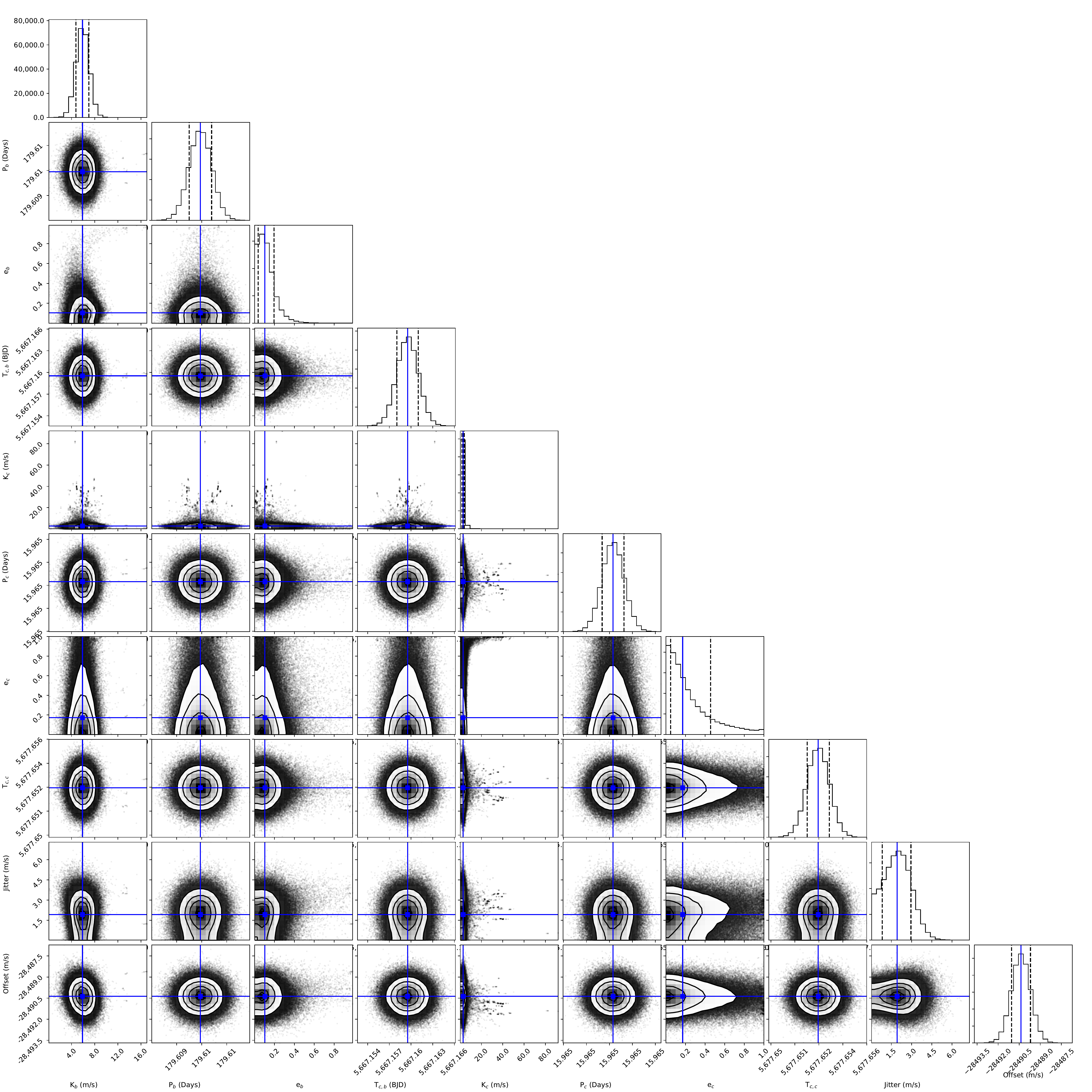}
    \caption{Posteriors distributions for the internal variables determined from the analysis of Kepler-103 RVs. The contours are at $1$, $2$, and $3 \sigma$. Median values for each parameter denoted by the solid blue lines are those given in Table \ref{tab:103_output}. Made using {\tt corner} \citep{corner}.}
	\label{fig:103_corners}
\end{figure*}

%%%%%%%%%%%%%%%%%%%%%%%%%%%%%%%%%%%%%%%%%%%%%%%%%%

% Don't change these lines
\bsp	% typesetting comment
\label{lastpage}
\end{document}